\documentclass{raa_rb}
\usepackage[a4paper]{geometry} 
\usepackage{graphicx,times}             
\usepackage{natbib}
\usepackage{amssymb,amsmath}
\usepackage{float}
\usepackage{subcaption}  
\usepackage{multirow}
\usepackage{booktabs}

\bibpunct{(}{)}{;}{a}{}{,}

\begin{document}

  \title{X-ray Sources Classification Using Machine Learning: A Study with EP-WXT Pathfinder LEIA
}

   \volnopage{Vol.0 (20xx) No.0, 000--000}      
   \setcounter{page}{1}          

   \author{Xiaoxiong Zuo 
      \inst{1,2,3}
   \and Yihan Tao 
      \inst{1,3,*}
   \and Yuan Liu
      \inst{1,*}
    \and Yunfei Xu
      \inst{1,2,3}
    \and Wenda Zhang
      \inst{1}
    \and Haiwu Pan
      \inst{1}
    \and Hui Sun
      \inst{1}
    \and Zhen Zhang
      \inst{1,2,3}
    \and Chenzhou Cui
      \inst{1,2,3}
    \and Weimin Yuan
      \inst{1}
   }

   \institute{National Astronomical Observatories, Chinese Academy of Sciences,
             Beijing 100101, China; 
        \and
             University of Chinese Academy of Sciences, Beijing 100049, China\\
        \and
             National Astronomical Data Center, 20A Datun Road, Beijing 100101, China\\
\vs\no
   {\small Received 20xx month day; accepted 20xx month day}}

\footnote{Corresponding author: y.tao@nao.cas.cn; liuyuan@nao.cas.cn}

\abstract{X-ray observations play a crucial role in time-domain astronomy. The Einstein Probe (EP), a recently launched X-ray astronomical satellite, emerges as a forefront player in the field of time-domain astronomy and high-energy astrophysics. With a focus on systematic surveys in the soft X-ray band, EP aims to discover high-energy transients and monitor variable sources in the universe. To achieve these objectives, a quick and reliable classification of observed sources is essential. In this study, we developed a machine learning classifier for autonomous source classification using data from the EP-WXT Pathfinder - Lobster Eye Imager for Astronomy (LEIA) and EP-WXT simulations. The proposed Random Forest classifier, built on selected features derived from light curves, energy spectra, and location information, achieves an accuracy of approximately 95\% on EP simulation data and 98\% on LEIA observational data. The classifier is integrated into the LEIA data processing pipeline, serving as a tool for manual validation and rapid classification during observations. This paper presents an efficient method for the classification of X-ray sources based on single observations, along with implications of most effective features for the task. This work facilitates rapid source classification for the EP mission and also provides valuable insights into feature selection and classification techniques for enhancing the efficiency and accuracy of X-ray source classification that can be adapted to other X-ray telescope data.
}


   \authorrunning{Zuo et al.}            
   \titlerunning{LEIA X-RAY SOURCES CLASSIFICATION USING ML}  

   \maketitle

%
%
\section{Introduction} \label{sec:intro}

X-ray observations are important sources for the study of time-domain astronomy. Transient sources within this field, such as supernovae, gamma-ray bursts (GRB), active galactic nuclei (AGN), and X-ray binaries (XRB), undergo substantial and sudden changes in radiation across the X-ray and gamma-ray spectra. Within the contemporary landscape of multiwave-length and multi-messenger time-domain astronomy, monitoring celestial events in the X-ray range holds great scientific promise. Notably, numerous X-ray satellites, such as Swift \citep{burrows2005swift}, XMM-Newton \citep{jansen2001xmm}, and Chandra \citep{weisskopf2000chandra}, are actively investigating the universe, producing a wealth of significant scientific discoveries. Time-domain phenomena are mostly characterized by their sporadic and transient nature. Rapid detection and timely monitoring of time-domain astronomical events are essential for their study and analysis. Efficient processing of large datasets is crucial in time-domain astronomy research, Furthermore, the integration of multi-wavelength data to reveal complex patterns and comprehensive understanding has resulted in a notable paradigm shift, with machine learning techniques emerging as prominent and influential tools.

Lobster Eye Micro-Pore Optics (MPO) is an innovative X-ray focusing technology known for its wide field of view and impressive imaging capabilities \citep{angel1979lobster,Hudec2010Kirkpatrick}. The Einstein Probe (EP), utilizing MPO technology, is a dedicated astronomical satellite designed for time-domain astronomy and high-energy astrophysics \citep{yuan2018einstein1}. Launched in January 2024, EP is equipped to perform rapid, high-frequency, and systematic surveys of the soft X-ray sky in the time-domain \citep{yuan2018einstein,yuan2015einstein}. The EP mission will enable rapid detection and precise localization of transient and variable sources, as well as the acquisition of high-quality light curves and spectral data. EP consists of 12 Wide-field X-ray Telescopes (WXT) covering the 0.5 to 4.0 keV range, accompanied by a Follow-up X-ray Telescope (FXT) that operates from 0.3 to 10 keV \citep{zhang2022estimate}. In July 2022, the Lobster Eye Imager for Astronomy (LEIA) was launched as the pathfinder for the WXT component of EP to verify its on-orbit performance and refine the operational parameters of the instrument. LEIA, equipped with a full-fledged WXT that offers an extensive field of view measuring 18.6°×18.6°, successfully completed its orbital tests \citep{zhang2022first}. It has obtained large-field X-ray measurement data for numerous celestial objects, revealing new transient sources \citep{yang2023leia,sun2023leia}, including the discovery of LXT 221107A \citep{ling2022leia, li2022swift}. The observational data gathered by LEIA establish a solid foundation and invaluable experience for the EP mission.

EP is positioned to collect a significant amount of time-domain sky survey data, primarily consisting of light curves and energy spectra. Employing artificial intelligence (AI) methodologies, including machine learning, to analyze extensive data resources of EP has the potential to uncover hidden insights in transient and variable sources. Researchers have developed a target detection framework using machine learning on the image data obtained by the Lobster Eye Telescope. This framework has been tested using EP-WXT simulation data, demonstrating promising accuracy and efficiency \citep{jia2023target}.

Since then, machine learning have attracted significant attention and achieved success in the automated classification of X-ray transient sources. This is evident from their application to data from XMM-Newton, Chandra, and other X-ray satellites. \cite{mcglynn2004automated} pioneered the application of machine learning techniques to classify X-ray sources, using oblique decision trees \citep{murthy1994system} to categorise approximately 80,000 sources from the ROSAT survey into six distinct categories : star, XRB, AGN, cluster, white dwarf and galaxies. However, the limited positional accuracy of ROSAT data, ranging from approximately 10'' to 30'', led to significant confusion in classification results. \cite{lo2014automatic} utilized a supervised learning approach to automatically classify 2XMMi-DR2 data obtained from the XMM-Newton mission. \cite{tranin2022probabilistic} applied naive Bayesian methods to Swift X-Ray Telescope (XRT) and XMM-Newton data. Additionally, \cite{zhang2021classification} cross-matched 4XMM-DR9, SDSS-DR12, and AIIWISE databases to extract multi-wavelength features for effective classification. \cite{yang2022classifying} employed Random Forest to classify Chandra Source Catalog version 2 (CSCv2) data, and developed MUWCLASS, an automated multi-band processing pipeline specifically designed for X-ray sources.

The application of AI technologies, particularly machine learning, in astronomical data classification can significantly reduce labor costs and improve classification efficiency. EP data presents unique challenges due to its short exposure times and limited photon counts of single observations, making source classificaiton and identification extremely difficult. Manual identification of the sources is labour-intensive and time-consuming, and may result in the unfortunate consequence of missing the optimal observational window for conducting follow-up observations.

The EP mission primarily aims to detect transient and variable sources. Due to the limited sampling points and photons in LEIA and EP data, and differences from X-ray telescopes such as Chandra, direct calculation of power-law distribution and periodic features is impractical. Consequently, the current X-ray source classification models and features developed for Chandra, XMM-NEWTON, and other data cannot be directly applied to EP data. EP is expected to accumulate a substantial amount of data; therefore, it is crucial to first conduct classification research on LEIA and EP data.  To support the requirements of EP team for single-observation classification and the discovery of new celestial objects, there is an urgent need for a machine learning classification algorithm capable of rapidly and accurately identifying transient and variable sources in real-time. 

In this paper, we propose a machine learning model that classifies target sources based on statistical characteristics of light curves, energy distributions, and other relevant features utilizing simulated EP data and LEIA observational data. The classification model has been implemented as a pipeline and deployed on the LEIA data processing server, enabling fast and real-time source classification during observations. The model can also be applied to EP in the future.

This paper is structured as follows. Section \ref{sec:Data} introduces the EP and LEIA data, presents the characteristics of simulated data, and describes the methods used for data pre-processing and dataset construction. Section \ref{subsec:feature} describes the feature extraction and selection for the classifier. Section \ref{sec:Method} provides the details of the processing methods and model optimization techniques used in this study. The performance of the developed model is presented in Section \ref{sec:5}. Section \ref{sec:6} discussed contentious issues in classification models, along with its application in the pipeline. Finally, Section \ref{sec:conclusion} provides a summary of the study, highlighting the key findings and potential implications for the field of astronomy.

\section{Data} \label{sec:Data}

\begin{figure}[htp]
    \centering
    \includegraphics[width=13cm]{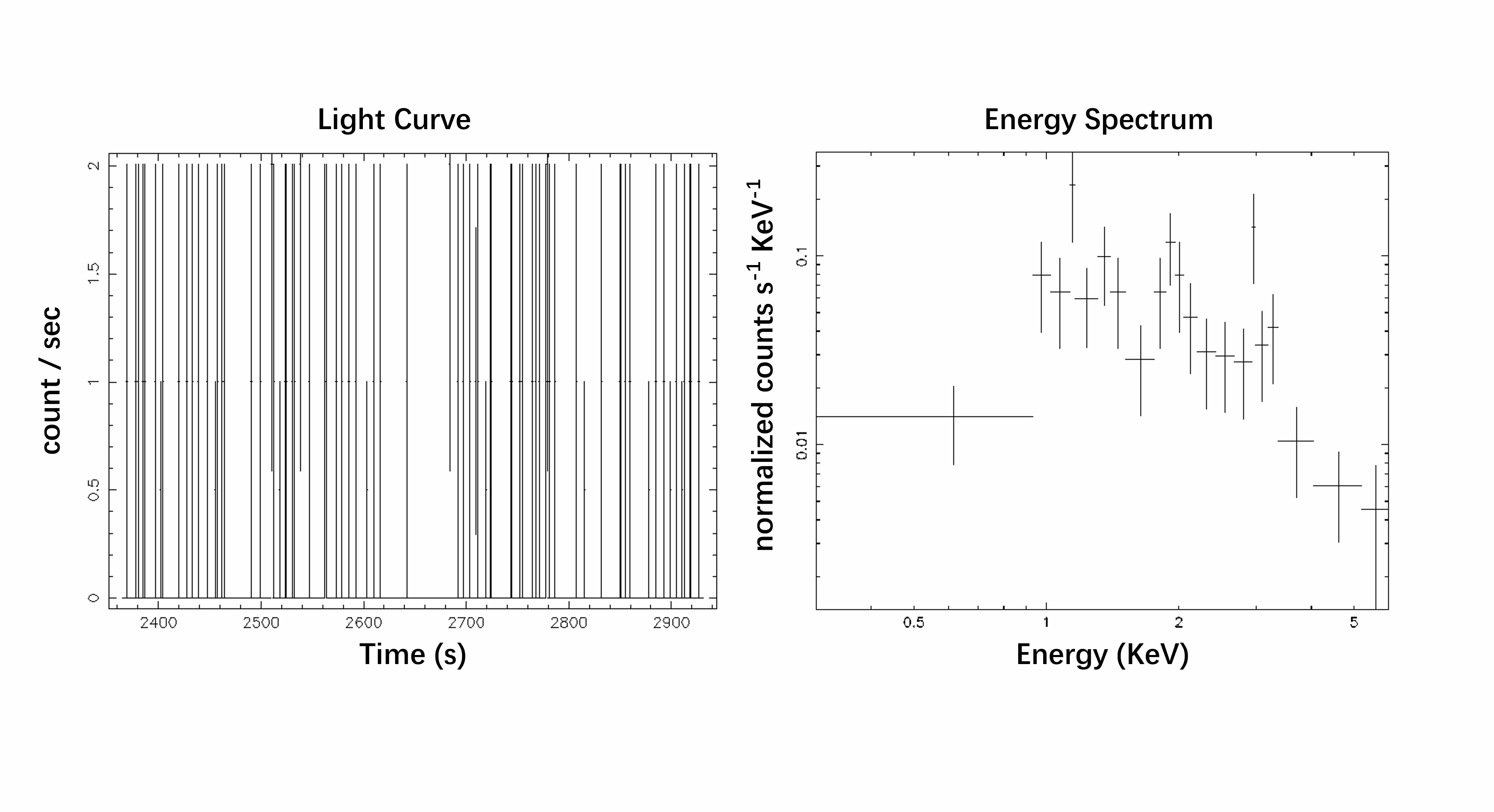}
    \caption{A quick view of an example of the LEIA data, where the left panel displays the light curve and the right panel displays the energy spectrum.}
    \label{fig:wxt13s2}
\end{figure}

The dataset utilized in this research comprises LEIA observational data and simulated EP data. The dataset is accessible within the China-VO PaperData Repository\footnote{https://doi.org/10.12149/101460}. Both the LEIA observational data and the simulated EP data encompass a variety of file types, including the catalog, spectrum, and light curve. The event file contains information about photon arrival times, photon energies, and the positional coordinates at which the photons intersect the detector plane. The catalog file serves as a high level data product of EP-WXT, containing information extracted from detected sources in a CMOS detector. This information includes counts, pixel positions, and celestial coordinates, and is stored in the catalog file as an index-ordered list of rows within a binary table extension. The light curve file is another high-level data product of EP-WXT generated from the event file of the EP pipeline. It provides the light curve for WXT, consisting of photon counting rates with a time resolution of 1 second. The spectrum files provide a record of the distribution of photon counts within the energy range of 0.5 to 4.0 keV. Figure \ref{fig:wxt13s2} displays the light curve and energy spectrum of a low mass X-ray binary as observed by the LEIA.

\subsection{LEIA data} \label{sec:Data1}

LEIA, as the pathfinder of EP-WXT, was responsible for carrying the complete WXT test module into orbit. LEIA has an 18.6°×18.6° field of view, an angular resolution of 3.8-7.5 arcmin, 4 CMOS sensors, a bandpass of 0.5-4.0 keV, an effective area of 2–3 $cm^2$ per 1 keV, a pixel size of 15µm, and a total of 4k × 4k pixels \citep{zhang2022first}.

\begin{figure}[htp]
    \centering
    \includegraphics[width=12cm]{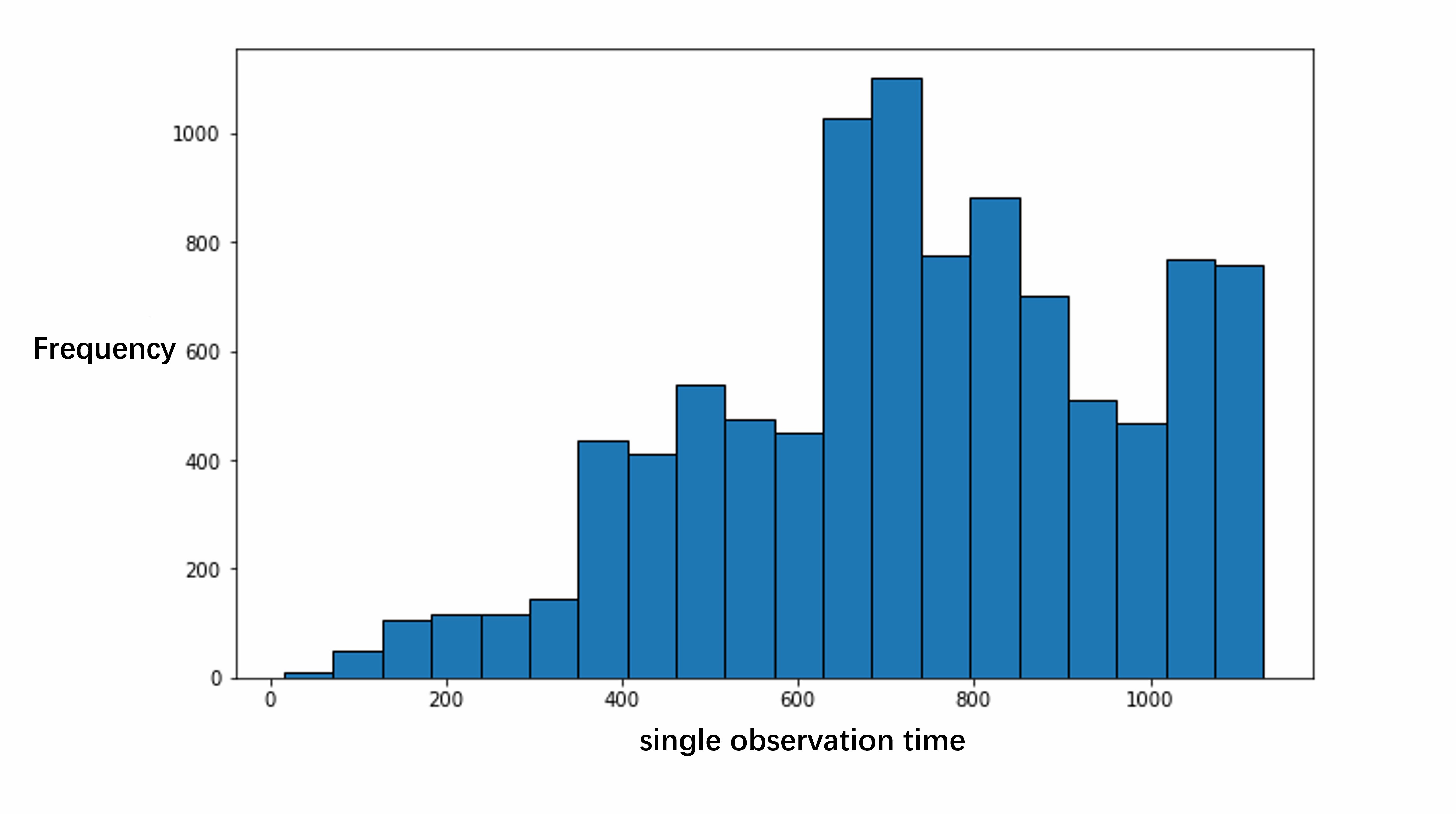}
    \caption{Histogram of LEIA single observation durations}
    \label{fig:distributionLIEA}
\end{figure}

By August 2023, LEIA had carried out 9,063 observations, with 8,172 of them undergoing manual verification. The data is obtained from the EP Time Domain Astronomical Information Center (TDIC)\footnote{https://ep.bao.ac.cn/leia/}, which provides functionalities for data querying and downloading. Artifacts were systematically removed from our dataset, resulting in categories such as AGN, XRB, stars, Galaxy Clusters, Pulsars, Supernova Remnants (SNR), and others. The data obtained by LEIA are derived from the certification conducted by the TA team after the operational activities of LEIA. The current known source categories were cross-validated with source tables from other satellites, allowing for the assignment of classification labels to the observed sources. Figure \ref{fig:distributionLIEA} illustrates the histogram of LEIA single observation durations, revealing that the data is relatively short, spanning from a few hundred seconds to over a thousand seconds. Table \ref{LEIA data} presents the quantity and proportion of each class of LEIA data.

\begin{table}
    \centering
    \caption{Class Distribution and Proportion in LEIA Data}
    \label{LEIA data}
    \begin{tabular}{ccc}
    \toprule
        Class & Number & Proportion  \\
    \midrule
        X-Ray Binary (XRB) & 4834 & 69.56\%  \\
        Supernova Remnant (SNR) & 949 & 13.66\%  \\
        Cosmic ray & 351 & 5.05\%  \\
        Active Galactic Nucleus (AGN) & 311 & 4.48\%  \\
        Star & 227 & 3.27\%  \\
        Pulsar & 159 & 2.29\%  \\
        Cluster of Galaxies & 118 & 1.70\%  \\
    \bottomrule
    \end{tabular}
\end{table}

\subsection{EP simulation data} \label{sec:Data2}

\begin{figure}[!ht]
    \centering
    \includegraphics[width=13cm]{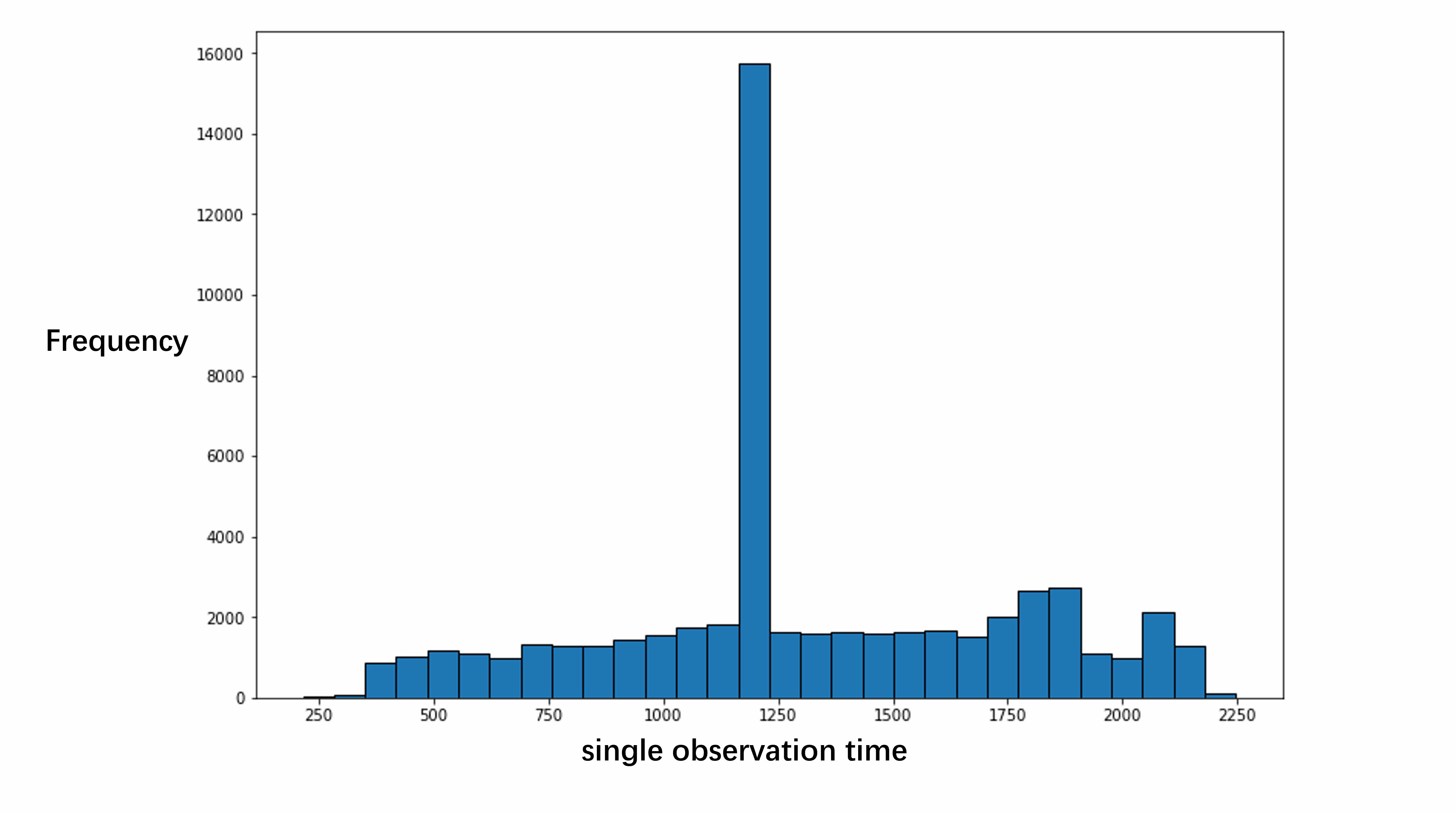}
    \caption{Histogram of EP simulation single observation durations}
    \label{fig:distributionEP}
\end{figure}

The EP simulation data were generated using a data simulator developed by the EP Science Application Team (Pan 2024, in preparation). The simulator employs the Monte Carlo method to generate data and introduces noise into the dataset. All target sources were selected from the ROSAT Skylight target directory. The simulation data are generated based on the pointing direction of EP, following the design specifications of the EP-WXT instrument. These simulated data closely replicate actual observational data and include event files, catalog files, spectrum files, light curve files, and other data types. The exposure times of simulated data range from 1,100 to 1,300 seconds, with a detector plane size of 4,096×4,096 pixels. The energy distribution of X-ray photons ranges from 0.5 to 4.0 keV. Figure \ref{fig:distributionEP} shows the histogram of EP simulation single observation durations, which primarily exceeds one thousand seconds.

The simulated data are categorized into 11 types: Active Galactic Nucleus (AGN), High-Luminosity Gamma-Ray Burst (HLGRB), Galactic Star, Galactic Compact Binary Black Hole, Galactic Compact Binary Neutron Star, Galactic Compact Binary Pulsar, Galactic Compact Binary White Dwarf, Galactic Compact Binary Neutron Star and Black Hole, Short-duration Gamma-Ray Burst (SGRB), Supernova Shock Breakout (SNSBO), and Tidal Disruption Event (TDE). To meet the requirements of the EP science team, and through simplification and consolidation, these categories were reclassified into seven types: AGN, Star, XRB, SGRB, HLGRB, SNSBO, and TDE.

The labels for the simulated data are derived from those assigned during the data generation process using the ROSAT catalog \citep{voges1999rosat}. In 1996, the X-ray source table from the ROSAT satellite sky survey was published, documenting over 18,000 X-ray sources with a positional accuracy of approximately 10 arcseconds.

As a single observation in the simulated data can capture multiple sources simultaneously or no source at all, we perform source location cross-matching between the catalog file and the simulated data directory. We use a matching radius of 0.05° and consider the matched data as a dataset for classification in our study. Due to the short exposure time during observations and instrumental limitations, there are a limited number of data points in the light curve. Therefore, light curve data with fewer than 350 data points were initially excluded. The classes of the simulated data are listed in Table \ref{EP data}.


\begin{table}
    \centering
    \caption{Class Distribution and Proportion in EP Simulation Data}
    \label{EP data}
    \begin{tabular}{cccc}
    \toprule
    Class                                     & \multicolumn{2}{c}{Number}     & Percentage               \\
    \cmidrule{2-3}
                                              & Count & Total Count &                          \\
    \midrule
    Galactic Compact Binary Neutron Star                & 10674 & \multirow{5}{*}{36803} & \multirow{5}{*}{60.74\%} \\
    
    Galactic Compact Binary White Dwarf                & 9242  &                        &                          \\
    Galactic Compact Binary Pulsar            & 8435  &                        &                          \\
    Galactic Compact Binary Black Hole                & 5452  &                        &                          \\
    Galactic Compact Binary Neutron Star and Black Hole            & 3000  &                        &                          \\
    Active Galactic Nucleus (AGN)             & \multicolumn{2}{c}{10662}     & 17.60\%                  \\
    Galactic star (Star)                      & \multicolumn{2}{c}{5496}      & 9.07\%                   \\
    Short-duration Gamma-Ray Burst (SGRB)     & \multicolumn{2}{c}{1575}      & 2.69\%                   \\
    Supernova Shock Breakout (SNSBO)        & \multicolumn{2}{c}{5559}      & 9.18\%                   \\
    Tidal Disruption Event (TDE)              & \multicolumn{2}{c}{376}       & 0.62\%                   \\
    High-Luminosity Gamma-Ray Burst (HLGRB)   & \multicolumn{2}{c}{117}       & 0.19\%                   \\
    \bottomrule
    \end{tabular}
\end{table}

\subsection{Training and test dataset} \label{dataset1}

Since the simulated data are generated based on the ROSAT star catalog, the available data categories are relatively limited. Consequently, categories such as cosmic rays and clusters of galaxies observed in LEIA are not included. However, it is important to note that LEIA has a relatively short operational period and has not yet observed rare categories such as GRBs and TDEs. These categories include rarely observed transient sources and unknown sources that are not currently present in the LEIA data but are the focus of EP's future detection efforts.

We have developed three approaches for constructing training sets for our model: using only EP simulated data, using only actual LEIA observational data, and combining EP simulated data with LEIA data. Section \ref{sec:5} provides a comprehensive description of the comparison among these three methods for constructing the datasets.

The final model was developed using a dataset that combines EP simulated data and LEIA observational data, encompassing all categories. The training dataset consists of EP simulated data and 80\% of the LEIA observational data, comprising a total of 32,406 data points. The distribution of classes in this merged dataset is shown in Table \ref{dataset}.

The data was divided in a class-balanced manner, with 80\% of the combined data used for model training and the remaining 20\% reserved as a mixed data test set to evaluate the model. Subsequently, the trained model is applied to the LEIA data. The remaining 20\% of LEIA observational data is designated as the LEIA data test set to assess the model's performance on the LEIA data. While the model was evaluated using both test sets, particular emphasis was placed on assessing its effectiveness on the LEIA data test set.

\begin{table}
    \centering
    \caption{Quantity of data on each class in different sets}
    \label{dataset}
    
    \begin{tabular}{cccc}
    \toprule
    Class                                     & Training Set & Mixed Data Test Set & LEIA Data Test Set  \\
    \midrule
    X-Ray Binary (XRB)                        & 16000        & 8141                & 985                 \\
    Active Galactic Nucleus (AGN)             & 8735         & 2182                & 56                  \\
    Star                      & 4534         & 1140                & 49                  \\
    Short-duration Gamma-Ray Burst (SGRB)     & 1269         & 306                 & 0                   \\
    Supernova Shock Breakout (SNSBO)        & 455          & 104                 & 0                   \\
    Tidal Disruption Event (TDE)              & 295          & 81                  & 0                   \\
    Supernova Remnant (SNR)                   & 604          & 154                 & 191                 \\
    High-Luminosity Gamma-Ray Burst (HLGRB)   & 92           & 25                  & 0                   \\
    Cosmic Ray                                & 237          & 51                  & 63                  \\
    Cluster of Galaxies                       & 74           & 21                  & 23                  \\
    Pulsar                                    & 111           & 25                  & 23                  \\
    Total                                     & 32406        & 12230                & 1390                \\
    \bottomrule
    \end{tabular}
\end{table}

As indicated in Table \ref{dataset}, our training dataset exhibits a significant label imbalance. The distribution of celestial body types across the celestial sphere is uneven, leading to a scarcity of certain rare transient sources. For instance, there is an overabundance of XRBs, while the number of rare sources such as GRBs and TDEs is inadequate. The number of XRB sources is approximately 300 times greater than that of HLGRBs. These imbalances can significantly impact the performance of machine learning algorithms. To address this issue, we utilized the Synthetic Minority Oversampling Technique (SMOTE; \citep{chawla2002smote}) to resample the dataset. This technique augmented the underrepresented categories and mitigated the problem of class imbalance. SMOTE employs the K-nearest neighbor method to generate synthetic samples for the minority class. This approach is commonly used to address class imbalance and is recognized for its robustness.

\begin{figure}[h!]
    \centering
    \includegraphics[width=12cm]{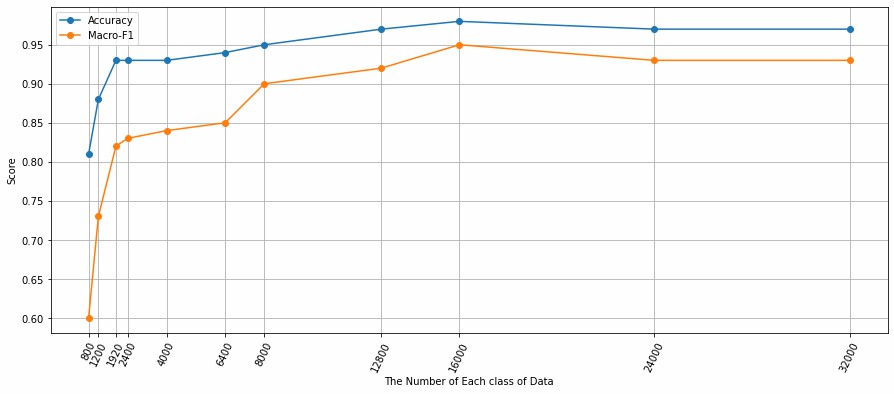}
    \caption{Accuracy and Macro-F1 under different sampling conditions of the data.}
    \label{fig:smote}
\end{figure}

We conducted experiments using various resampling scenarios and observed that increasing the volume of data led to improved outcomes. We attempted undersampling the categories with sufficient data, such as XRB and AGN, while resampling the remaining classes. The impact of different sample sizes, ranging from 800 to 32,000, is illustrated in Figure \ref{fig:smote}. We found that once the data quantity exceeded 16,000, there were diminishing returns in terms of accuracy and Macro-F1 scores. Consequently, our final approach involved applying SMOTE to resample all classes except for XRB. For XRB, we performed random undersampling to obtain 16,000 samples. This resulted in a total of 16,000 samples for each class.

Due to certain limitations in feature calculation, the value of some feature may be null or infinite. In such cases, we assign a uniform value of -100 to these features. Filling in missing values in this manner does not impact the model's performance.

\section{Feature extraction}\label{subsec:feature}

The short time scale of a single EP observation poses a challenge in capturing the periodic behaviour of the targets in the time domain. Additionally, instrumental limitations constrain the applicability of commonly used features, such as time variability, periodicity, power law, and flare-related features, to EP data. The extraction of features that uncover the underlying physical significance of the data is a critical task. The process of feature extraction necessitates meticulous consideration of their underlying physical meanings. The design of the feature extraction method refers to the studies conducted by \cite{lo2014automatic} and \cite{richards2011machine}. 
Table \ref{tab:astro-phenomena} summarizes the key characteristics of the different astronomical transients and variables considered in this study, including their timescales, light curve characteristics, and energy spectrum characteristics, while the energy spectrum for each class can be found in Appendix \ref{app:spectrum} and the light curves in Appendix \ref{app:lightcurves}. Table \ref{feature} presents a list of the 23 features derived from the data, along with their detailed descriptions. The final selection includes the top 9 features. This section delves into the design and extraction of features.

\begin{table}[htbp]
    \hfill
    \centering
    \caption{Characteristics of Sources and Phenomena in X-ray Band}
    \label{tab:astro-phenomena}
    \resizebox{\textwidth}{!}{%
    \begin{tabular}{cccc}
        \toprule
        Class & Time Scale & Light Curve Characteristics & Energy Spectrum Characteristics \\
        \midrule
        AGN & Minutes to years & Aperiodic variability & Low flux density \\
        XRB & Milliseconds to years & Strong aperiodic variability & High flux density \\
        Cluster of Galaxies & - & No variability & Low flux density \\
        Cosmic Ray & Milliseconds & Photons concentrated in a single readout frame & Very low flux density \\
        Pulsar & Milliseconds to years & Periodic and aperiodic variability & High flux density \\
        SNR & - & No variability & High flux density \\
        Star & Kilo-seconds to years & Weak variability; occasionally exhibiting a stellar flare & Low flux density \\
        TDE & Minutes to years & Weak variability & Low flux density, Soft spectrum \\
        SNSBO & Minutes to hours & Transient flare & Low flux density, Soft spectrum \\
        GRB & Seconds to minutes & Transient short-term flare & High flux density, Hard spectrum \\
        \bottomrule
    \end{tabular}
    }
    \hfill
\end{table}

\begin{table}[!ht]
    \centering
    \caption{List of Time Series Features Used for Classification. \\The top 9 features are selected features.}
    \label{feature}
    \resizebox{\textwidth}{!}{%
    \begin{tabular}{lll}
        \toprule
        Feature & Description & References \\
        \midrule
        galactic longitude & Galactic longitude of source  &\quad  \\
        galactic latitude & Galactic latitude of source  &\quad  \\
        a\_hard & The count rates in the 0.2 – 0.5 keV  &\quad  \\
        b\_hard & The count rates in the 0.5 – 1.0 keV  &\quad  \\
        c\_hard & The count rates in the 1.0 – 2.0 keV  &\quad  \\
        de\_hard & The count rates above 2.0 keV  &\quad  \\
        kurt & Kurtosis of the distribution of count rates; calculated using scipy.stats.kurtosis  &\quad  \\
        skew & Skewness of the distribution of count rates; calculated using scipy.stats.skew & \citet{lo2014automatic,richards2011machine} \\
        modulation index & Variance / mean & Improvement from \citep{lo2014automatic}\\
        var & Variance of the counts &\quad  \\
        beyond1Std & Percentage of observations that lie beyond one standard deviation from the mean & \citet{lo2014automatic,richards2011machine} \\
        energy\_ratio & The ratio between peak energy and background energy  &\quad  \\
        mean & Mean of the counts  &\quad \\
        median & Median of the counts  &\quad  \\
        percentile\_diff & Count rate at the 98th percentile minus the count rate at the 2nd percentile& \citet{lo2014automatic,richards2011machine} \\
        maximum slope & Maximum slope of adjacent observation points & \citet{lo2014automatic,richards2011machine}\\
        median\_abs\_deviation & Median of the absolute value of the deviation from the median & \citet{lo2014automatic,richards2011machine} \\
        percentage\_within\_threshold & Percentage of measurements within 20\% of the median & \citet{lo2014automatic,richards2011machine}\\
        t50 & 50\% energy width on the sides of the peak position of the energy spectrum &\quad  \\
        t20 & 20\% energy width on the sides of the peak position of the energy spectrum &\quad  \\
        t10 & 10\% energy width on the sides of the peak position of the energy spectrum &\quad  \\
        t50\_t20 & t50 / t20 &\quad  \\
        t50\_t10 & t50 / t10 & \quad\\
        \bottomrule
    \end{tabular}%
    }
\end{table}

\subsection{Spectral features}

The energy spectrum range detected by the EP spans from 0.5 to 4.0 keV. The hardness ratio is a widely employed feature in X-ray astronomy for characterizing the spectral morphology of X-ray sources. This method involves comparing the photon counts detected in two or more distinct energy bands, typically categorizing them into high-energy (hard X-rays) and low-energy (soft X-rays) bands. Although the hardness ratio is a fundamental and effective technique, it may not fully capture the complexities of an X-ray spectrum.

Hardness ratios are determined by analyzing the counts in selected energy bands, which are chosen based on the specific instrument and the scientific questions under investigation. In this study, we have delineated the following energy bands: 0-0.5 keV, 0.5-1.0 keV, 1.0-2.0 keV, and above 2.0 keV. Experimental comparisons indicate that utilizing energy band counting yields superior results.

Energy band counting involves partitioning the X-ray data into distinct energy ranges and enumerating the photon counts detected in each band. This approach provides more granular information, allowing for a more comprehensive understanding of the characteristics of X-ray sources. Consequently, energy band counting enhances the robustness of the analytical algorithms employed. The energy spectrum for each class can be seen in the Appendix \ref{app:spectrum}.

\subsection{Light curve feature}

The power-law characteristics cannot be investigated due to the brevity of the light curve. However, specific statistical features can be extracted from the available data. Below are some of the features that are essential for model training. The light curve for each class can be seen in the Appendix \ref{app:lightcurves}.

\subsubsection{kurt}

Kurtosis is a statistical feature used to describe the distribution of a light curve. It quantifies the sharpness or flatness of a probability distribution curve relative to its mean. More specifically, kurtosis characterizes the steepness of the data distribution curve relative to the standard normal distribution. For our calculations, we utilize the \textit{scipy.stats.kurtosis} function from the SciPy package \citep{2020SciPy-NMeth}.

Kurt is defined by the following equation:
\begin{equation}
    \label{kurt}
    kurt = \frac{{m}_{4}}{{m}^{2}_{2}}
\end{equation}
${m}{2}$ represents the second-order central moment (variance) of the dataset, and ${m}{4}$ represents the fourth-order central moment of the dataset.

\subsubsection{Skew}

Skewness is a statistical measure used to quantify the asymmetry of a probability distribution. In light curves, skewness indicates the degree of asymmetry in the temporal changes of luminosity. A positive skewness value suggests a longer right tail in the distribution, while a negative skewness value implies a longer left tail. In astronomy, various celestial objects exhibit diverse patterns of luminosity changes. Quantifying the skewness of a light curve enables us to understand the probability distribution characteristics of observed luminosity changes, thereby enhancing our comprehension of the underlying physical processes. For calculations, we employ the \textit{scipy.stats.skew} function from the SciPy package \citep{2020SciPy-NMeth}.

Skewness is defined as follows:
\begin{equation}
    \label{skew}
    Skew = \frac{1}{n} \sum_{i=1}^{n} \frac{{(x_i - \overline{x})^3}}{s^3} 
\end{equation}
${n}$ is the number of data points and ${s}$ is the standard deviation.


\subsubsection{Modulation Index}

The relative volatility index can be obtained by dividing the variance of the light curve by its mean. This characteristic is referred to as the \textquoteleft modulation index\textquoteright. The relative volatility allows for the comparison of volatility among different light curves. A large relative volatility indicates significant changes in the photometric data, while a small relative volatility suggests relatively stable changes. A higher relative volatility value indicates a more active and unstable light variation phenomenon.

\begin{figure}[htbp]
  \centering
  \begin{minipage}[b]{0.48\textwidth}
    \centering
    \includegraphics[width=\textwidth]{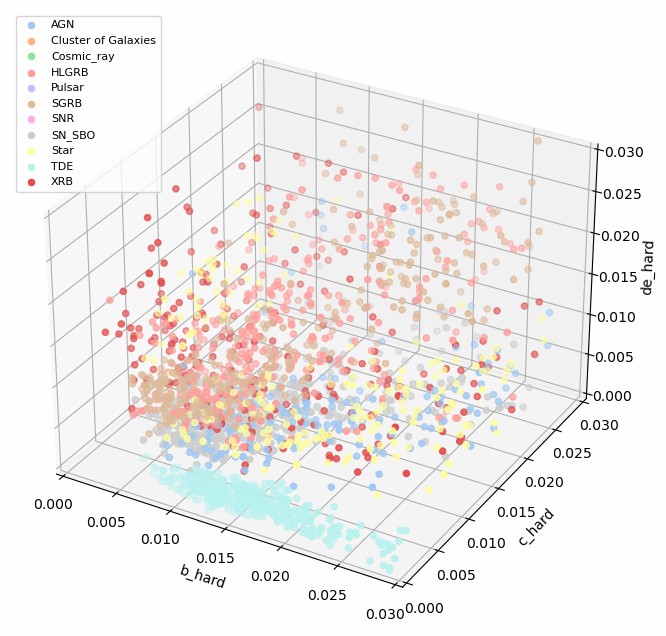}
    \caption{The distribution of each class of data in b\_hard, c\_hard and de\_hard three features}
    \label{fig:3dplot}
  \end{minipage}%
  \hfill%
  \begin{minipage}[b]{0.48\textwidth}
    \centering
    \includegraphics[width=\textwidth]{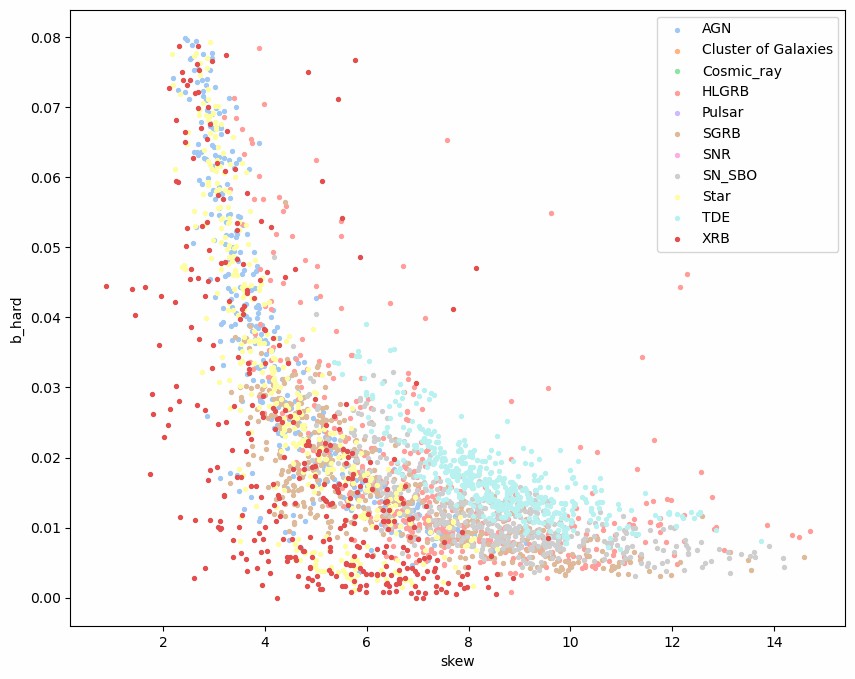}
    \caption{The distribution of data on the two features of skew and b\_hard}
    \label{fig:skew_vs_b_hard}
  \end{minipage}
\end{figure}

We randomly selected 500 data points from each class to examine their distribution in the feature space using different features, as depicted in Figure \ref{fig:3dplot} and Figure \ref{fig:skew_vs_b_hard}. It is evident that these features effectively distinguish between the classes. Figure \ref{fig:3dplot} illustrates the distribution of data from different classes in three hardness ratio spaces: 0.5-1.0 keV, 1.0-2.0 keV, and above 2.0 keV. Figure \ref{fig:skew_vs_b_hard} presents the distribution of data from different classes based on skewness and the 0.5-1.0 keV hardness ratio.

\subsection{The distribution of the data on the sky map}

The types of X-ray sources can be partially distinguished by considering Galactic longitude and Galactic latitude. Previous studies have utilized Galactic latitude as a classification feature \citep{lo2014automatic,tranin2022probabilistic,mcglynn2004automated}. The distribution of Galactic longitude and Galactic latitude is also influenced by the telescope's survey design. In our study, we consider the use of Galactic longitude and Galactic latitude as effective features for classifying sources in the EP data.

\begin{figure}[ht]
    \centering
    \includegraphics[width=14cm]{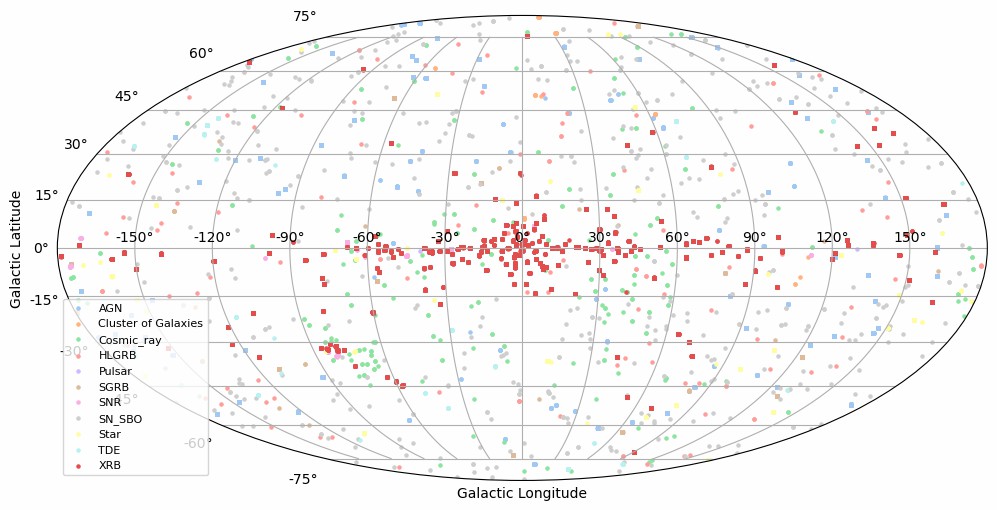}
    \caption{The distribution of the data on the sky map}
    \label{location}
\end{figure}

Machine learning algorithms that incorporate Galactic longitude and Galactic latitude as features of position information can achieve higher accuracy rates due to the distinct position distributions of various celestial objects. By combining Galactic longitude and Galactic latitude as location information features with other attributes of celestial objects, more complex feature vectors can be constructed. Currently, the manual determination of the EP observation source also takes into account the location information for assessment. EP observations yield a relatively large number of high-energy celestial objects, including XRBs, which tend to cluster near the Galactic center. Figure \ref{location} illustrates the spatial distribution of the data in the observed sky area.

\section{Classification Methods and Procedures}\label{sec:Method}

All sources are labeled. We extract above described features from the data of observation sources with labels for supervised learning. This section will introduce the process of data processing, feature extraction, and machine learning in detail.

\subsection{Algorithm} \label{subsec:Algorithm}

In this work, our primary algorithm of choice is Random Forest, an ensemble learning technique that harnesses the collective power of multiple decision trees \citep{breiman2001random}. The core principle underlying Random Forest revolves around its combination of bootstrap aggregating and random feature selection. By employing these strategies, the algorithm aims to introduce increased randomness and diversity into the model, thereby enhancing its overall performance.

The strength of Random Forest lies in its ability to generate an ensemble prediction by aggregating the outputs of all the individual trees through either majority voting for classification tasks or averaging for regression tasks. The collective decision-making process of Random Forest leads to robust predictions that exhibit high accuracy and stability.

Random Forests are particularly well-suited for handling datasets with a large number of input features and samples. The algorithm's inherent randomness and the aggregation of multiple trees allow it to effectively capture complex relationships and patterns within the data. Furthermore, Random Forests offer the advantage of assessing feature importance. By analyzing the contribution of each feature to the model's performance, we gain valuable insights into the key factors driving the observed patterns and outcomes.

We employed the Random Forest algorithm to learn and harness the light curve variability features and spectral features of the data, achieving remarkable results. By leveraging the power of ensemble learning, Random Forest effectively captured the intricate patterns and relationships within the data. The fusion of the data's temporal dynamics and spectral attributes within the Random Forest framework proved highly effective, with the algorithm's ability to combine the predictive strengths of multiple decision trees through voting yielding impressive outcomes.

\subsection{Feature selection}

Feature importance quantifies the impact of individual features on the performance of a machine learning model. This analysis aids in identifying the most influential features, thereby enhancing model efficiency, interpretability, and our understanding of the factors driving the predictions.

In order to identify the most essential features, we initially incorporated 23 features for training the classifier. Table \ref{feature} provides a comprehensive list of these features along with their descriptions. We evaluated the contributions of features across various samples, analyzing both their individual and cumulative significance.

\begin{figure}[ht]
    \centering
    \includegraphics[width=14cm]{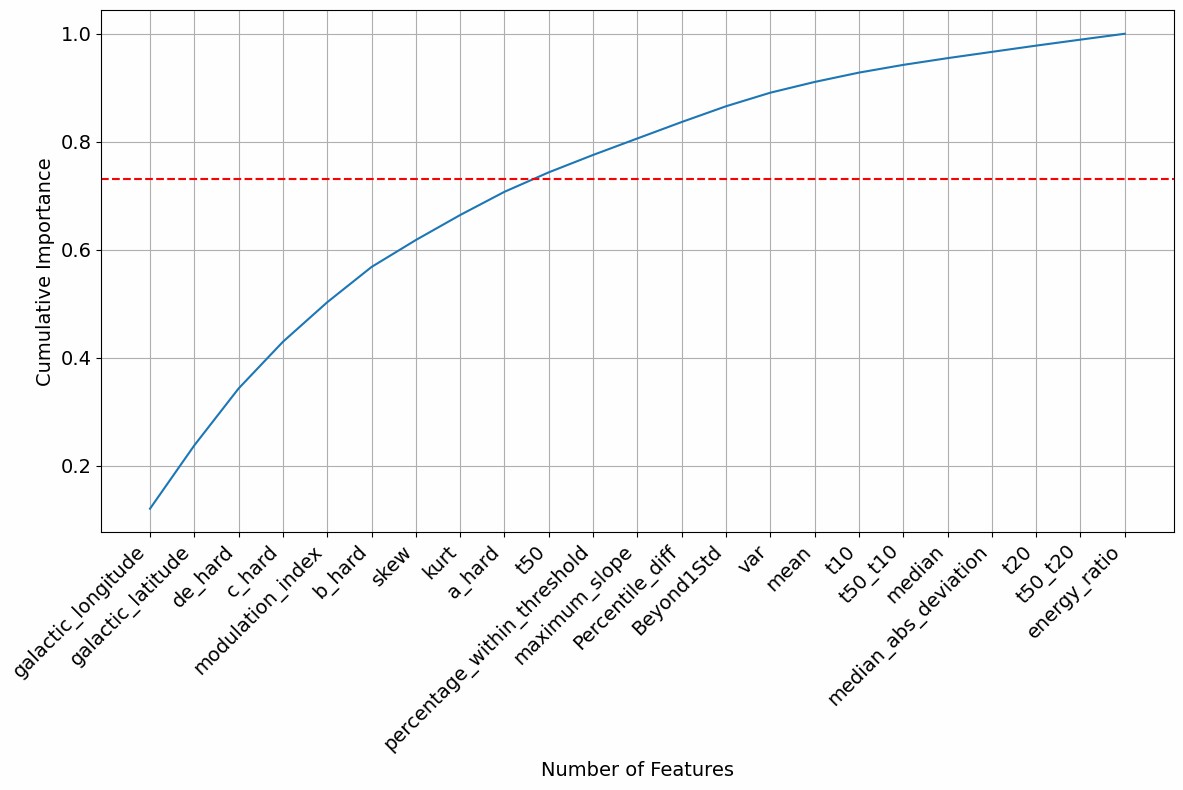}
    \caption{The cumulative importance of features. The red line is a threshold of 0.8.}
    \label{importance}
\end{figure}

Due to the characteristics of the data, not all features provide equal informativeness when applied to EP data. Features with lower importance are deemed to have limited significance. To select the most informative features, we applied a threshold based on the cumulative importance score. Figure \ref{importance} displays the cumulative importance ranking of all features, and we selected features with a cumulative importance score below 0.73.

When utilizing all 23 features, the classification performance was good, but after performing feature selection and reducing the feature set to 9, the classifier's performance improved significantly in both the mixed data test set and LEIA data test set. The classification accuracy and Macro-F1 score increased notably in both test sets. Table \ref{feature selection} presents the results of the feature selection comparison, highlighting the improved performance achieved by utilizing the selected subset of features.

\begin{table}
\centering
\caption{Comparison of the effect of feature selection}
\label{feature selection}
\begin{tabular}{cccccc}
\toprule
& \multicolumn{2}{c}{LEIA data test set} & \multicolumn{2}{c}{mixed data test set} \\
\cmidrule(lr){2-3}\cmidrule(lr){4-5}
& Accuracy & Macro-F1 & Accuracy & Macro-F1 \\
\midrule
23 features & 96.8\% & 92.3\% & 88.0\% & 83.2\% \\
9 selected features & 97.8\% & 94.4\% & 95.0\% & 85.4\% \\
\bottomrule
\end{tabular}
\end{table}

\subsection{Cross-Validation}

Cross-validation is a statistical technique used to assess the performance and generalization ability of machine learning models. It involves dividing the dataset into multiple subsets, iteratively training the model on a portion of the data, and validating it on the remaining subsets. This process is repeated multiple times, with different subsets used for training and validation in each iteration.

By employing cross-validation, we can obtain more robust and reliable performance evaluation results, as the model is tested on multiple subsets of the data rather than relying on a single train-test split. This approach helps to mitigate overfitting and provides a more accurate estimate of the model's performance on unseen data. The results of the cross-validation are presented in Table \ref{cv}.

\begin{table}[!ht]
    \centering
    \caption{Cross validation results}
    \label{cv}
    \begin{tabular}{lll}
    \toprule
        & Average & Variance  \\
    \midrule
        Accuracy & 98.5\% & 6.199e-07  \\
        Macro-F1 & 87.0\% & 7.824e-05   \\
    \bottomrule
    \end{tabular}
\end{table}

\subsection{Hyperparameter selection}

We use cross-validated grid search to select the best hyperparameters. Grid search evaluates the performance of each parameter combination by searching for the best parameter combination in the parameter space and using cross validation. In the cross validation process, the dataset is divided into 5 folds, with 1 fold used as the validation set and the other 4 folds used as the training set each time. This can comprehensively evaluate the performance of the model and reduce the impact caused by the randomness of dataset partitioning.

\begin{figure}[htbp]
  \centering
  \begin{minipage}[t]{0.48\textwidth}
    \centering
    \includegraphics[width=\textwidth]{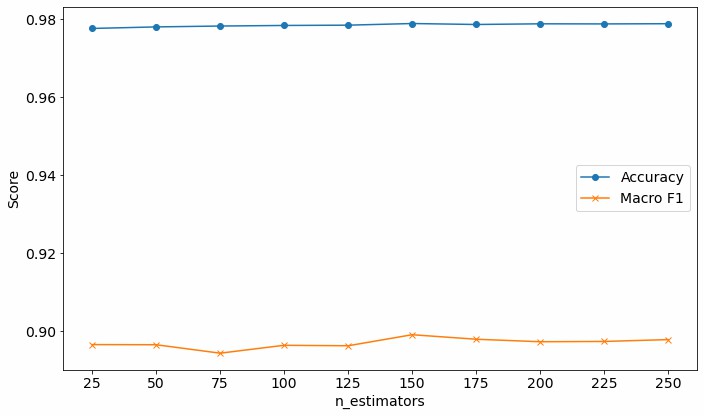}
    \label{fig:es}
  \end{minipage}
  \hfill
  \begin{minipage}[t]{0.48\textwidth}
    \centering
    \includegraphics[width=\textwidth]{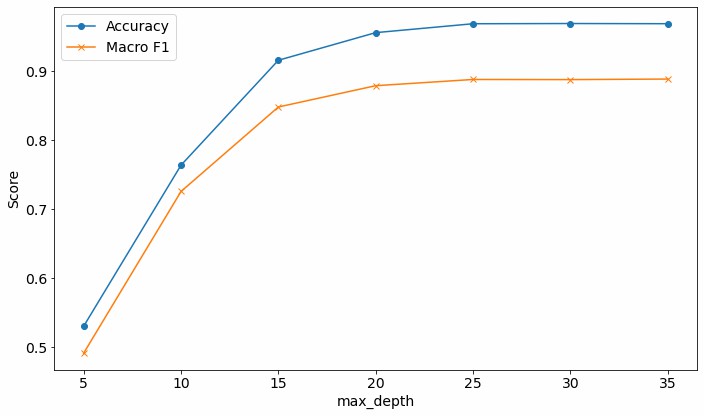}

    \label{fig:de}
  \end{minipage}
  \caption{Figure of hyperparameter selection. The left figure displays accuracy and Macro-F1 scores for different n\_estimators values, while the right figure illustrates accuracy and Macro-F1 scores for different max\_depth values.}
  \label{fig:hy}
\end{figure}

The Random Forest algorithm consists of three main adjustable hyperparameters: the total number of trees ($n\_estimators$), the maximum depth of each decision tree ($max\_depth$), and the maximum number of features used by each tree node ($max\_features$). We choose the default $'auto'$ value for $max\_features$. Using cross-validated grid search, we evaluated the hyperparameters $n\_estimators$ and $max\_depth$ while keeping the other hyperparameters fixed in their default settings. 
The results of the hyperparameter selection are illustrated in Figure \ref{fig:hy}. Our findings indicate that the model performs best when hyperparameters are set to $n\_estimators = 
 150$ and $max\_depth = 25$.

\section{Results} \label{sec:5}

\subsection{Evaluation indicators}\label{Evaluation indicators}

In this study, we employ five evaluation metrics to assess the effectiveness of the classification models. These metrics serve as robust indicators of model performance, including accuracy, balanced accuracy, Macro-F1 score, Matthews Correlation Coefficient (MCC), and run time.

Accuracy is calculated by dividing the number of correctly classified samples by the total number of samples.


Balanced Accuracy is calculated by averaging the accuracies of each class, resulting in a balanced accuracy indicator that effectively addresses the bias caused by data imbalance. The formula for calculating Balanced Accuracy is
\begin{equation}
    \label{BAccuracy}
    Balanced Accuracy = \frac{1}{N} \sum_{i=1}^{N} \frac{TP_i}{P_i}
\end{equation}
Among them, ${N}$ is the number of classes, ${TP}_{i}$ is the true number of samples in the ${i}$-th class, and ${P}_{i}$ is the total number of samples in the ${i}$-th calss.

The Macro F1 score is an evaluation metric that considers both accuracy and recall. It is calculated by averaging the precision and recall values across all categories, resulting in the Macro-F1 value. This metric treats each class equally, making it robust to data imbalance.





The Matthews correlation coefficient (MCC) is an evaluation indicator that provides a comprehensive measure of the relationship between true positive, true negative, false positive, and false negative predictions. It is particularly suitable for datasets with imbalanced categories. The MCC value ranges from -1 to 1, where 1 indicates perfect prediction, 0 represents random prediction, and -1 indicates completely inconsistent prediction.

\begin{equation}
    \label{MCC}
   MCC = \frac{TP \times TN - FP \times FN}{\sqrt{(TP + FP)(TP + FN)(TN + FP)(TN + FN)}}
\end{equation}

Run time is a practical metric that measures the time required for model training. It reflects the efficiency and speed of the model, making it particularly valuable when dealing with large datasets.

\subsection{Algorithms Comparison}\label{algo1com}

In this research paper, we attempted several popular machine learning algorithms.

XGBoost \citep{chen2016xgboost} is a gradient boosting framework that integrates regularization and parallel processing. Compared to Random Forests, XGBoost's strategy is more focused and sequential. For our comparative study, we implemented XGBoost using the Python library $xgboost$, aligning hyperparameters with those of the Random Forest model to ensure a fair comparison. The results showed that XGBoost's performance was competitive, nearly matching the robustness of Random Forest predictions.

The K-Nearest Neighbors (KNN) algorithm \citep{cover1967nearest}  classifies data through the majority vote of its k nearest neighbors in the feature space. For our analysis, we used the $KNeighborsClassifier$  from the sklearn library  \citep{pedregosa2011scikit}, setting n\_neighbors=5 as the default. The algorithm uses the Minkowski distance metric and a leaf node size of 30 to balance efficiency and accuracy. The uniform weighting scheme ensures equal contribution from all neighbors.

The Naive Bayes classifier is a probabilistic model that applies Bayes' theorem while assuming feature independence. We implemented the $GaussianNB$ from  Sklearn. Despite the model's simplicity, its assumptions of feature independence and Gaussian distribution can be restrictive, potentially affecting its performance in complex datasets where these conditions are not met. However, its effectiveness in the probabilistic classification of X-ray sources as studied by \cite{tranin2022probabilistic}, highlights its utility in specific applications.

Support Vector Machines (SVMs) \citep{cortes1995support} are a class of powerful supervised learning models known for their ability to find an optimal hyperplane that separates different classes with the maximum margin. In our predictive model, we employed the sklearn.svm library, opting for the RBF kernel. This approach, while effective, comes with challenges such as increased sparsity in high-dimensional spaces and sensitivity to feature selection. The computational demands of SVMs grow with the number of features, leading to longer training times and higher memory consumption.

We employed various machine learning algorithms for data classification. The evaluation metrics for assessing the model include accuracy, balanced accuracy, Macro-F1, MCC, runtime. Through comprehensive comparison and evaluation, we determined that Random Forests exhibit superior performance. The performance comparison among different algorithms is presented in Table \ref{diff}.

\begin{table}[!ht]
    \centering
    \caption{Performance Comparison of Different Algorithms}
    \label{diff}
    \begin{tabular}{cccccc}
    \toprule
        ~ & RandomForest & XGBOOST & KNN & GaussianNB & SVM  \\ \midrule
        Accuracy & 97.8\% & 96.5\% & 91.4\% & 28.1\% & 39.2\%  \\
        Balanced accuracy & 95.5\% & 95.2\% & 91.5\% & 51.9\% & 60.0\%  \\
        Macro-F1 & 94.3\% & 81.5\% & 64.2\% & 30.5\% & 30.1\%  \\
        MCC & 95.5\% & 93.2\% & 84.6\% & 18.1\% & 36.9\%  \\
        Run Time & 84.902 & 127.022 & 0.283 & 0.048 & 1087.391  \\ \bottomrule
    \end{tabular}
\end{table}

\subsection{The Final Pipeline Performance Evaluation}

\begin{figure}[ht!]
    \centering
    \includegraphics[width=12cm]{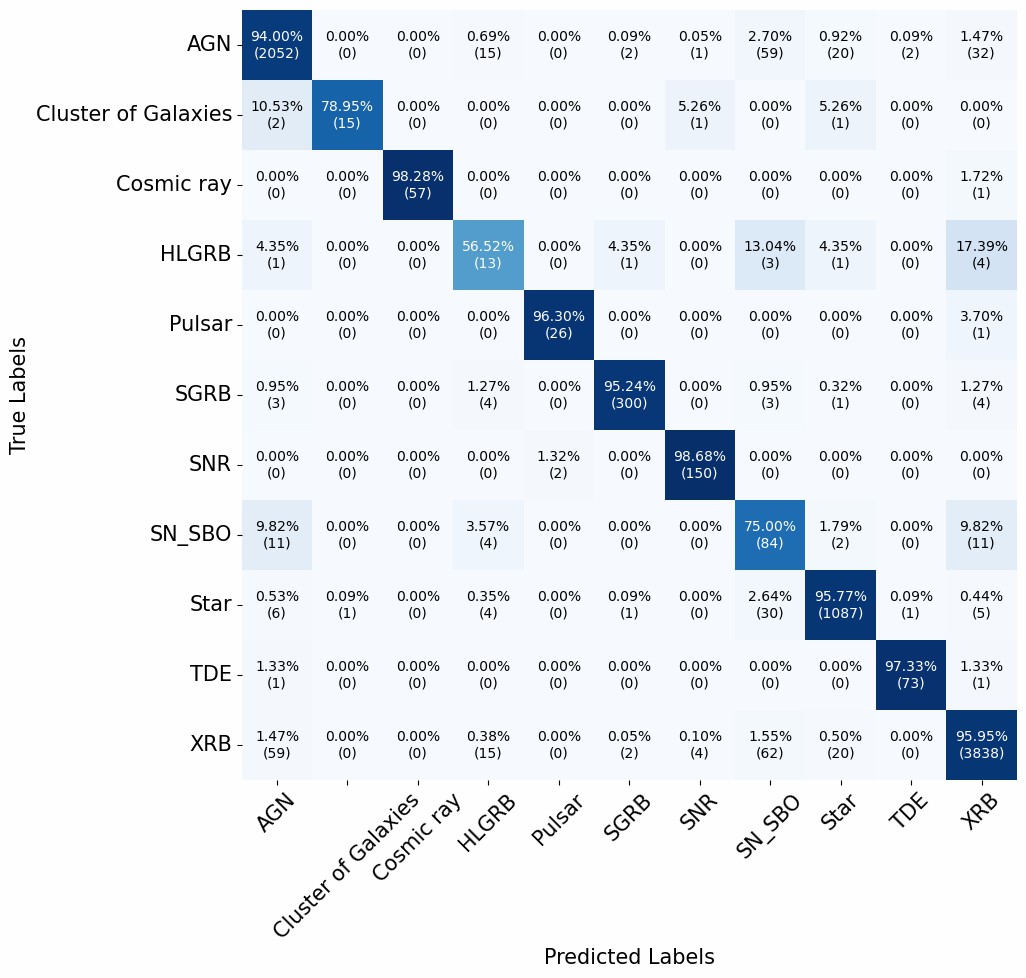}
    \caption{Confusion Matrix on Mixed Data Test Set.}
    \label{cm1}
\end{figure}

\begin{figure}[ht!]
    \centering
    \includegraphics[width=12cm]{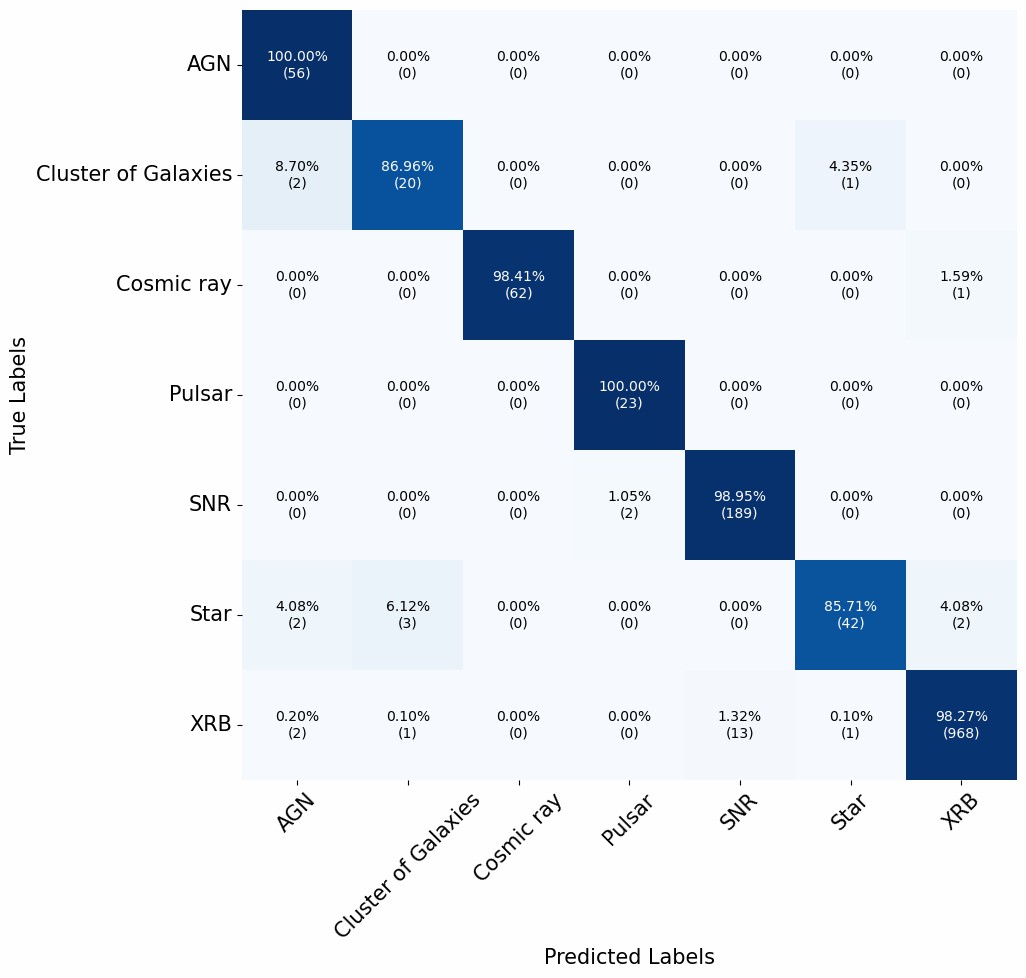}
    \caption{Confusion Matrix on the LEIA Data Test Set.}
    \label{cm2}
\end{figure}

Finally, we conducted experiments using the Random Forest algorithm with the hyperparameters \textit{n\_estimators} = 150 and \textit{max\_depth} = 25, while utilizing 9 feature parameters for classification. The mixed data was employed as the final training set. The accuracy achieved on the mixed data test set is 95.0\%, while the accuracy on the LEIA test set is 97.8\%. The specific details of the final test results are presented in Table \ref{performance}. Figures \ref{cm1} and \ref{cm2} display the confusion matrices for the LEIA test set and the mixed data test set.

\begin{table}[!ht]
    \centering
    \caption{Performance Evaluation of the Final Pipeline}
    \label{performance}
    \begin{tabular}{ccccc}
    \toprule
        Data & Accuracy & Balanced accuracy & Macro-F1 & MCC  \\ \midrule
        Mixed data test set & 95.0\% & 89.2\% & 85.4\% & 92.6\%  \\
        LEIA data test set & 97.8\% & 95.5\% & 94.3\% & 95.5\%  \\ \bottomrule
    \end{tabular}
\end{table}

In the mixed data test set, the classification results remain unsatisfactory for certain classes, such as HLGRB, SNSBO, and others, which comprise a small number of rare time-domain targets. This is primarily due to the limited number of objects in these classes. The LEIA survey has not yet identified these rare time-domain objects, and the data used continues to be simulated data from the ROSAT star catalog. In contrast, cosmic ray targets lack a light curve, allowing for their successful identification based on other characteristic features. The accuracy rate for identifying cosmic ray targets reaches 100\%.

\section{Discussion and Application} \label{sec:6}

\begin{figure}[ht!]
    \centering
    \includegraphics[width=12cm]{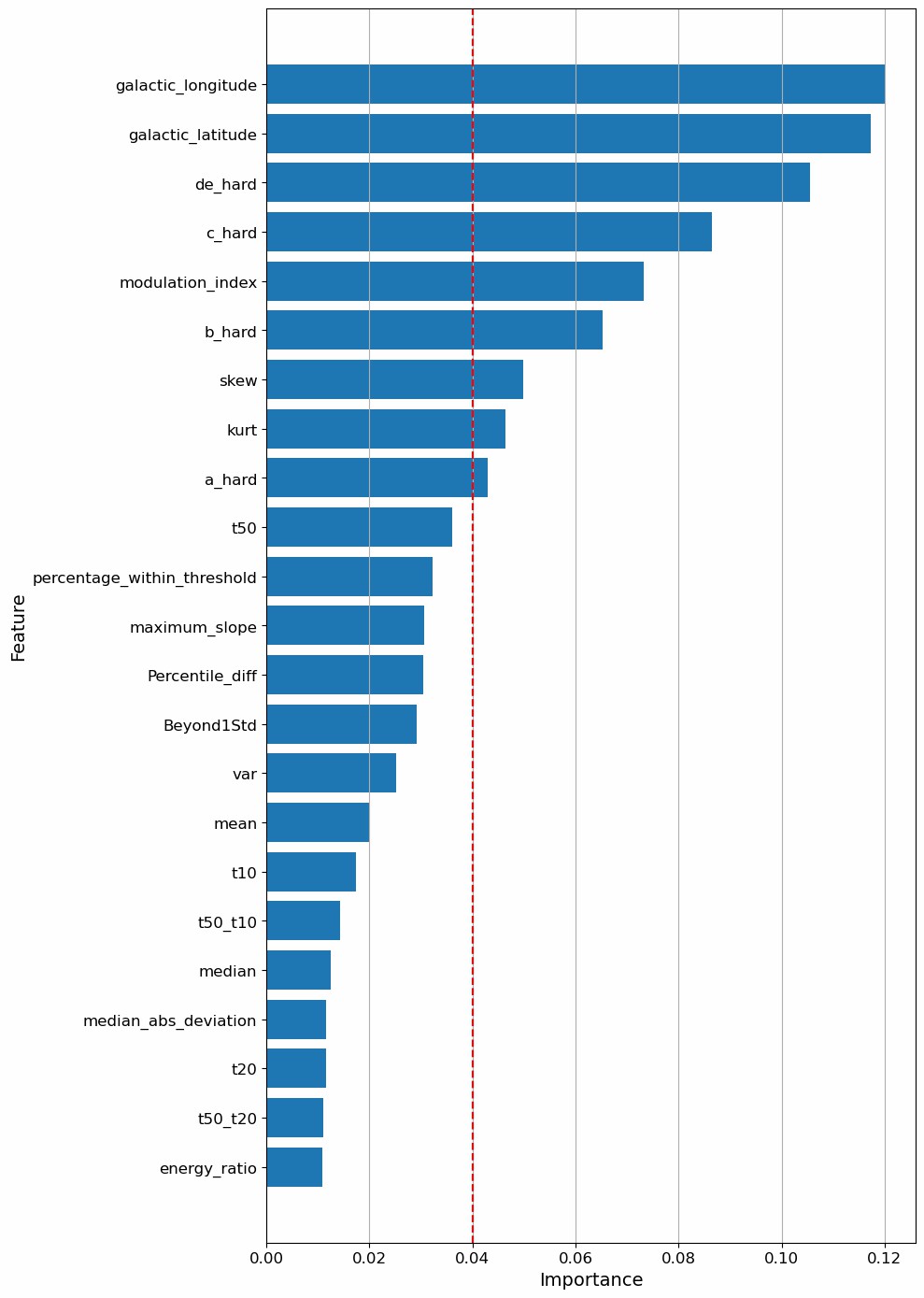}
    \caption{Feature Importance Ranking}
    \label{feaimp}
\end{figure}

Random Forests construct each decision tree by randomly selecting a subset of features, which effectively mitigates the risk of overfitting in high-dimensional spaces. Each tree is trained on a distinct subset of features, thereby reducing the model's dependence on any individual feature. Such diversity significantly enhances the model's generalization ability. Selecting appropriate features for modeling in high-dimensional spaces can be challenging, but Random Forests excel at handling numerous features without requiring explicit feature selection. The model can automatically identify significant features from the entire set and maintain relatively high performance, even in the presence of irrelevant features. Figure \ref{feaimp} illustrates the ranking of feature importance, highlighting the top 9 features of significant importance.

\subsection{Interpretability and feature importance}\label{sec:51}

Figure \ref{feature} demonstrates the innovative aspects of our feature design. Extracting features from the light curve data in previous studies, such as power-law fitting and Lomb-Scargle periodogram, has proven highly challenging. Three distinct features within the light curve — \textquoteleft kurt\textquoteright, \textquoteleft skew\textquoteright, and \textquoteleft  modulation index\textquoteright— were identified as significant features. These three attributes capture the sharpness, skewness, and oscillatory behavior manifested in the light curve, respectively. The classification of cosmic rays and stars relies significantly on these light curve features. Cosmic rays have almost no light curve characteristics, only photons concentrated in a single readout frame. Stars have weak variability, occasionally exhibiting a stellar flare. 

Regarding spectral features, we opted not to utilize the hardness ratio method, which may not capture all the characteristics of the X-ray spectrum. Instead, we enhanced previous research by replacing the hardness ratio with energy band counting, leading to improved outcomes. Our study revealed that features associated with energy spectra distribution, specifically \textquoteleft b\_hard\textquoteright \ and \textquoteleft c\_hard\textquoteright, exhibited remarkable significance in the model. This can be attributed to the primary role of EP as an X-ray telescope, specializing in the study of high-energy celestial objects. Three distinct features are capable of capturing the energy spectral distribution across the soft to hard X-ray energy bands for different classes. Therefore, incorporating energy spectra distribution as features provides substantial advantages in this specific context. These features make significant contributions by providing valuable information and aiding in the classification process. For instance, \textquoteleft b\_hard\textquoteright\ and \textquoteleft de\_hard\textquoteright\ made valuable contributions to classifying SNR, XRB, and cluster of galaxies, effectively distinguishing or excluding these classes and thereby improving the screening of AGN, among others.  

Additionally, utilizing \textquoteleft galactic longitude\textquoteright\ and \textquoteleft galactic latitude\textquoteright\ provides an efficient means of classifying galactic sources, including XRB and pulsars. Table \ref{tab:high-contribution-features} describes the high contribution features for different classes in LEIA data.

\begin{table}[h]
\centering
\caption{High Contribution Features for Different Classes in LEIA Data}
\label{tab:high-contribution-features}
    \begin{tabular}{cc}
        \toprule
            Class & High Contribution Features \\
            \midrule
            AGN & a\_hard, kurt, galactic longitude, galactic latitude \\
            XRB & galactic longitude, galactic latitude, de\_hard, c\_hard \\
            Cluster of Galaxies & galactic longitude, galactic latitude, a\_hard, b\_hard \\
            Cosmic Ray & kurt, skew, modulation index \\
            Pulsar & galactic longitude, galactic latitude, b\_hard, c\_hard \\
            SNR & galactic longitude, galactic latitude, b\_hard \\
            Star & kurt, skew, galactic latitude, c\_hard \\
            \bottomrule
    \end{tabular}
\end{table}

\begin{figure}[ht!]
    \centering
   
    \includegraphics[width=14cm]{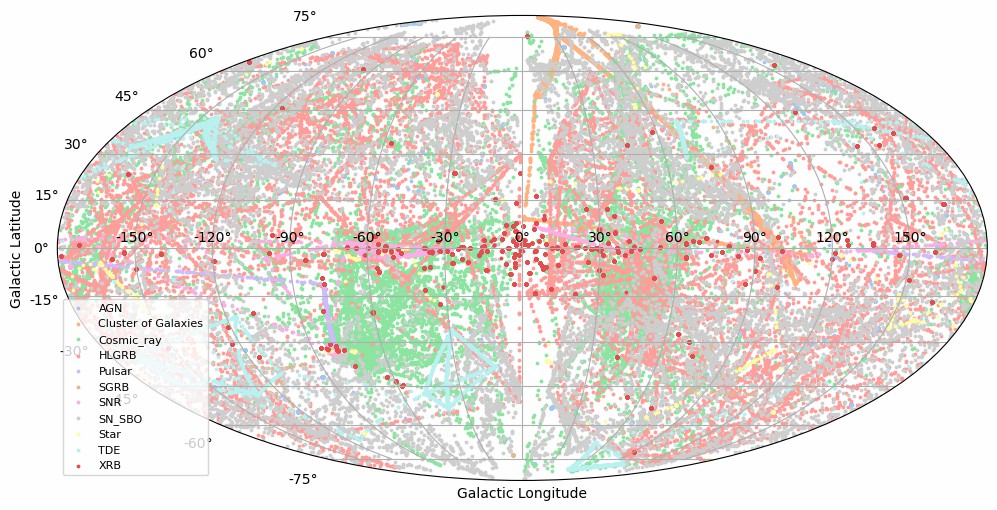}
    \caption{The distribution of the data on the sky map after SMOTE resampling}
    \label{loca2}
\end{figure}

Simulated data frequently include repeated observations of the same celestial sources, resulting in the presence of multiple observations for individual sources. We recognized that directly utilizing spatial information could lead to excessive overfitting of sources with multiple observations within the spatial feature space. To address this concern, we employed a resampling technique, specifically SMOTE, on the \textquoteleft galactic longitude\textquoteright \ and \textquoteleft galactic latitude\textquoteright\. Consequently, the spatial positions of the sources displayed a randomized distribution across the sky map, as illustrated in Figure \ref{loca2}.

\subsection{Comparison of different methods for constructing the training set}\label{52}

We assessed the effectiveness of three dataset construction methods: utilizing only EP simulation data, only LEIA data, and merging both EP and LEIA data. The evaluation was conducted using a consistent set of features. Given the model's intended use as a pipeline for classification tasks in both LEIA and EP, we measured the efficacy of these three training set construction methods using untrained LEIA data as the test set.

In the instance where simulated data served as the direct training set, SMOTE resampling was applied uniformly to each class to achieve 16,000 data points per class and a Random Forest classifier was utilized. Notably, the simulated data lacked certain categories detected by LEIA, such as SNR, cosmic rays, and clusters of galaxies. Consequently, the categories common to both datasets were restricted to XRB, AGN, and stars, exclusively reserved for testing in the LEIA test set.

When employing actual LEIA observational data as the training set, the SMOTE resampling strategy was adjusted to accommodate the smaller dataset size, resulting in 3,387 samples per class. However, the test set featured fewer categories due to the unique presence of classes in the simulated data not observed by LEIA. Classification testing was carried out using a consistent approach. It is essential to highlight that in scenarios with limited data volume, there is a risk of overfitting post-resampling. Detailed outcomes for these three scenarios are outlined in Table \ref{datacom123}.

\begin{table}[h!]
    \centering
        \caption{Comparison of the effects of three methods of constructing training sets}
    \label{datacom123}
    \begin{tabular}{lllll}
    \toprule
        Training set & Accuracy & Balanced accuracy & Macro-f1 & MCC  \\ \midrule
        Only Simulate data & 90.8\% & 28.5\% & 47.4\% & 36.3\%  \\
        Only LEIA data & 98.2\% & 97.0\% & 96.0\% & 96.4\%  \\
        Mixed data & 97.8\% & 95.5\% & 94.3\% & 95.5\%  \\ \bottomrule
    \end{tabular}
\end{table}

Given the current operational timeframe, data scarcity, and observation region constraints, it is crucial to acknowledge these limitations despite the promising results derived from utilizing LEIA data for training. Additionally, LEIA data covers a narrower spectrum of source categories. Notably, anticipated future surveys by both LEIA and EP are projected to detect specific time-domain targets like SGRB and SNSBO, present in simulated data but not yet observed. Classifying these targets will aid in identifying novel celestial objects. Thus, we have opted to continue incorporating EP simulated data alongside LEIA data in our training model.

\subsection{Pipeline application}

\begin{figure}[ht!]
    \centering
    \makebox[\textwidth]{\includegraphics[width=1.2\textwidth]{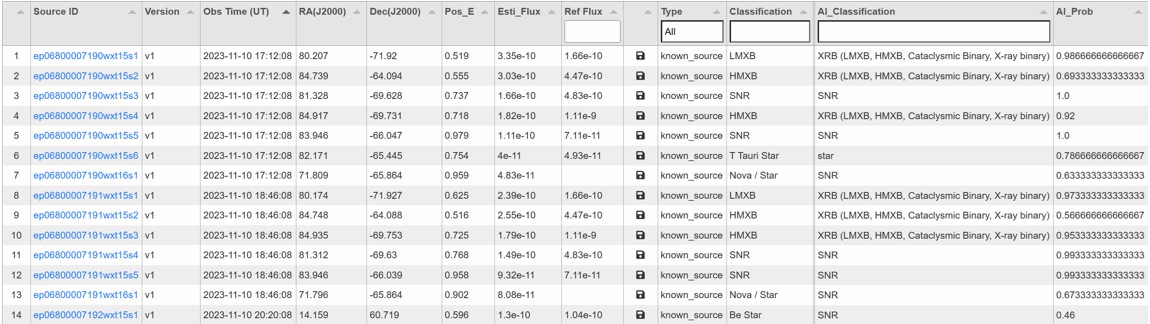}}

    \caption{Automatic classification of observation sources in the EP data processing interface}
    \label{TA}
\end{figure}

\begin{figure}[ht!]
    \centering
    \includegraphics[width=13cm]{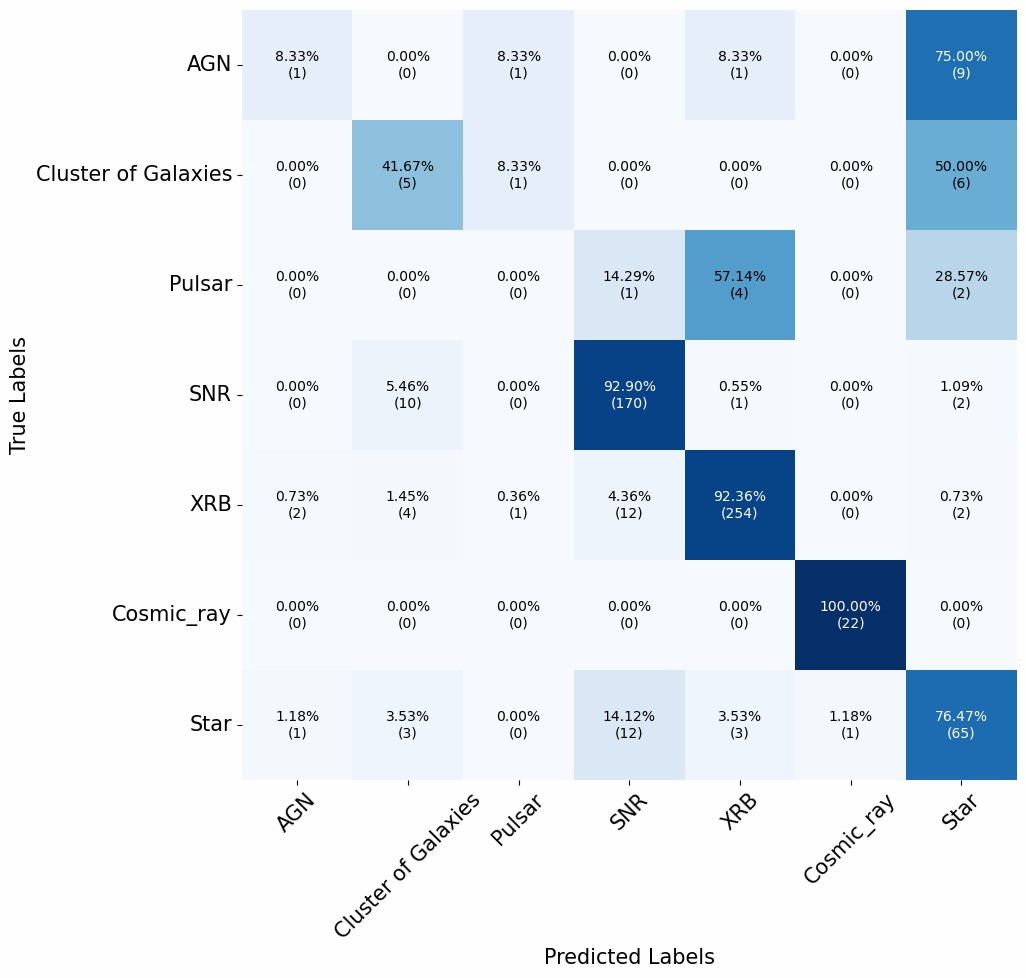}

    \caption{Confusion Matrix for Classification of Models in LIEA Observational Data in October and Novemberin 2023}
    \label{cmleia}
\end{figure}

The trained classification model has been encapsulated in a Docker contrainer and integrated into the data processing pipeline. The classification model delivers AI-based classification outcomes for each observation, along with the corresponding probability of the predicted class. The processing time for a single observation within the pipeline is approximately 0.19 seconds, while an observation may encompass multiple sources. This capability significantly aids the Transient Advocate team in validating observed sources. Figure \ref{TA} showcases the automated classification of observation sources within the EP data processing interface. Actual observation data from LEIA in October and November was selected to evaluate the classification results of the application model. After filtering out interference items like arm and fake sources from this dataset, a total of 596 instances were analyzed. The classification accuracy for this data stands at 86.7\%, with the classification confusion matrix depicted in Figure \ref{cmleia}.

Upon examination, it was noted that misclassifications within the AGN category were prevalent across multiple observations from three sources, primarily being classified as stars. AGN and stars share similar physical characteristics, leading to potential confusion. In the case of Pulsars, misclassifications were observed in multiple observations from two sources. One of them, the Magnetar, was misclassified as a star or an SNR, largely due to physical resemblance. The other source, a Pulsar, was classified as an XRB, primarily influenced by its proximity to the galactic center. The galactic location erroneously attributed to XRB based on feature contributions. Lastly, the classification performance for clusters of galaxies is suboptimal due to the limited representation of galaxy clusters in the training set, impacting the model's overall performance.

\subsection{Limitation}

The dataset includes a limited number of instances belonging to the cluster of galaxies and HLGRB. Consequently, these minority class instances may face challenges during SMOTE resampling, increasing the risk of overfitting. This risk arises from the potential generation of synthetic samples that closely mimic the minority class instances, potentially exacerbating overfitting concerns. Additionally, the application of SMOTE can lead to increased class overlap, blurring the distinctions between classes and impacting the model's decision boundary. This could make it more difficult to differentiate between different categories. For example, the resampling process had a notable impact on the positional distribution of the Cluster of Galaxies, as illustrated in Figure \ref{loca2}. The availability of more extensive data in the future is expected to alleviate this limitation.

The features of \textquoteleft galactic longitude\textquoteright \ and \textquoteleft galactic latitude\textquoteright \ are indicative of the distribution of sources on the sky map. They are particularly effective in classifying sources that are in the galactic center. However, when dealing with sources that are situated at significant distances from the galactic center and possess high galactic latitudes, such as high-latitude XRB and SNR,  these features may exert an inverse effect within the classifier, potentially leading to misclassification.

The subpar classification performance also can be attributed to frequent observations from diverse sources, leading to inadvertent classification errors. Notably, a significant portion of misclassifications exhibit a classification probability below 0.5, typically hovering around 0.3 or 0.4. To mitigate this issue, we plan to introduce a probability threshold as a filter in our forthcoming work to enhance the classification accuracy.

\section{Conclusion} \label{sec:conclusion}

The paper primarily delves into investigating a time-domain target classification algorithm tailored for X-ray telescopes. This research is executed through empirical analysis utilizing simulated data from EP and observational data from LEIA. A dataset is curated by combining EP simulation data with LEIA measurements, and a distinct set of classification features tailored for X-ray telescope data is proposed. This approach showcases promising performance in scenarios characterized by limited data points and shorter observation durations. Following a comparative analysis of various machine learning algorithms, Random Forest is selected as the classification algorithm, achieving an accuracy rate of 97.8\%. Moreover, this study integrates classification models into data processing pipelines, facilitating classification predictions for newly detected sources. The implications of this research are notably significant for the data processing tasks associated with EP missions. Upon EP's acquisition of fresh data, the classification model will be leveraged to classify categories not previously observed by LEIA. The findings presented in this paper can serve as a valuable resource for data analysis in high-energy space satellite missions and time-domain astronomy.

%

\begin{acknowledgements}
This work is supported by the National Key Research and Development Program of China (2022YFF0711500), National Natural Science Foundation of China (NSFC)(12373110, 12273077, 12103070, 12333004), and the Strategic Priority Research Program of the Chinese Academy of Sciences Grant No.XDA15310300, Grant No.XDB0550200, Grant No. XDB0550100, and Grant No.XDB0550000. Data resources are supported by China National Astronomical Data Center (NADC) and Chinese Virtual Observatory (China-VO). This work is supported by Astronomical Big Data Joint Research Center, co-founded by National Astronomical Observatories, Chinese Academy of Sciences and Alibaba Cloud.

\end{acknowledgements}

\appendix                  

\clearpage
\section{The energy spectrum for each class}\label{app:spectrum}

\begin{figure}[h!]
    \centering
    \begin{minipage}{0.5\textwidth}
        \includegraphics[width=\linewidth]{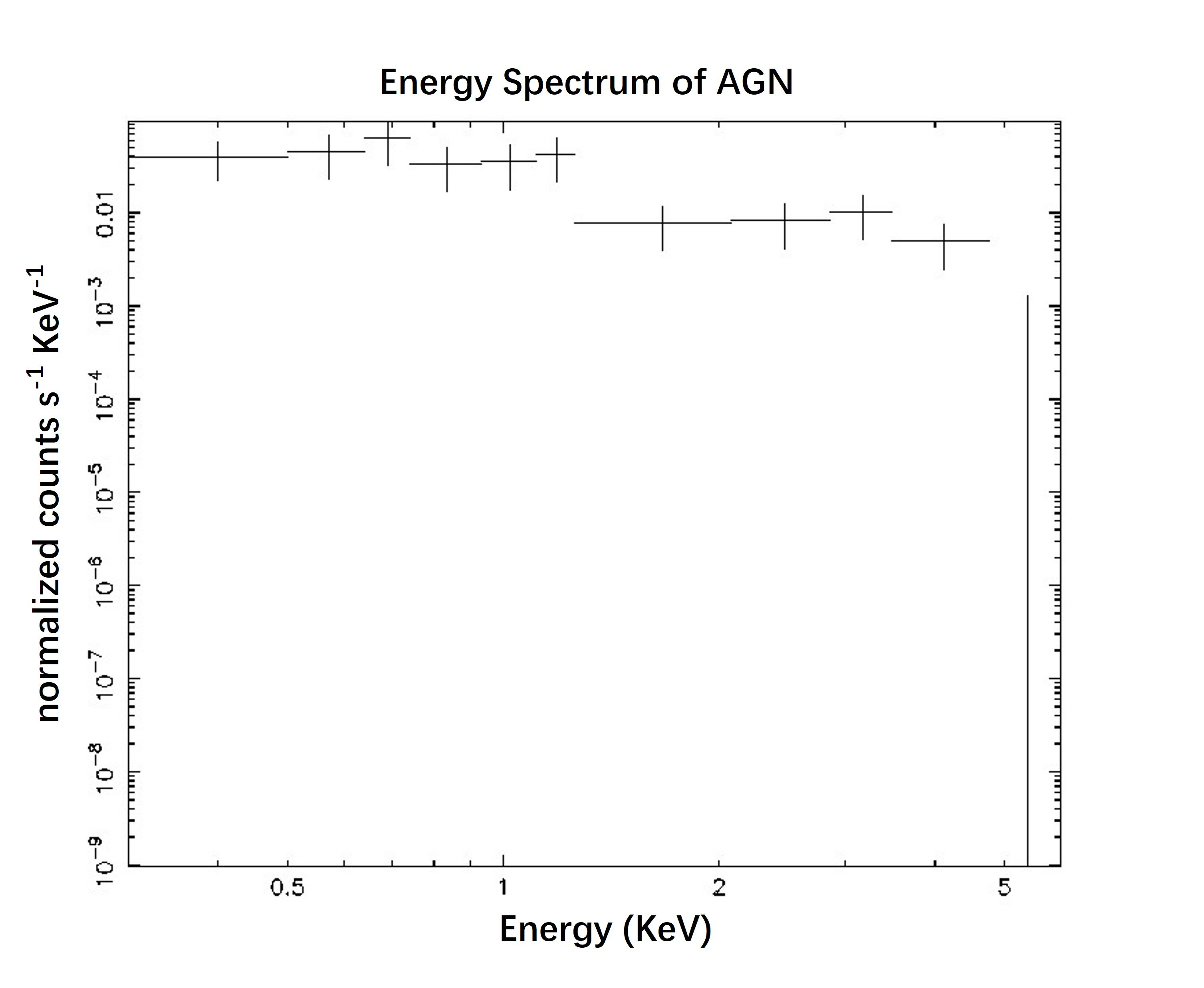}
        \caption{Energy spectrum of AGN}
        \label{BL Lac AGN LC}
    \end{minipage}%
    \begin{minipage}{0.5\textwidth}
        \includegraphics[width=\linewidth]{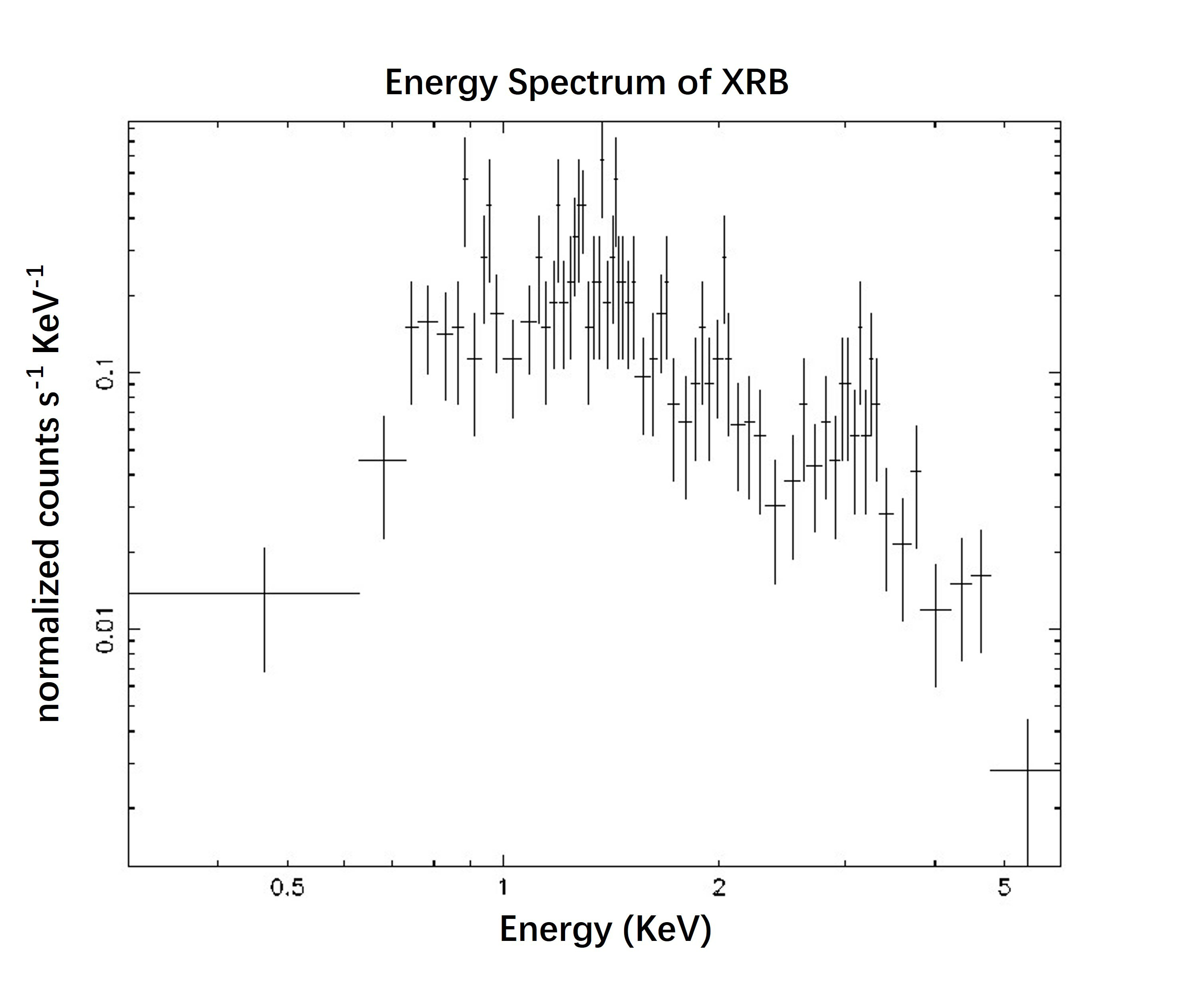}
        \caption{Energy spectrum of XRB}
        \label{LMXB LC}
    \end{minipage}
\end{figure}

\begin{figure}[h!]
    \centering
    \begin{minipage}{0.5\textwidth}
        \includegraphics[width=\linewidth]{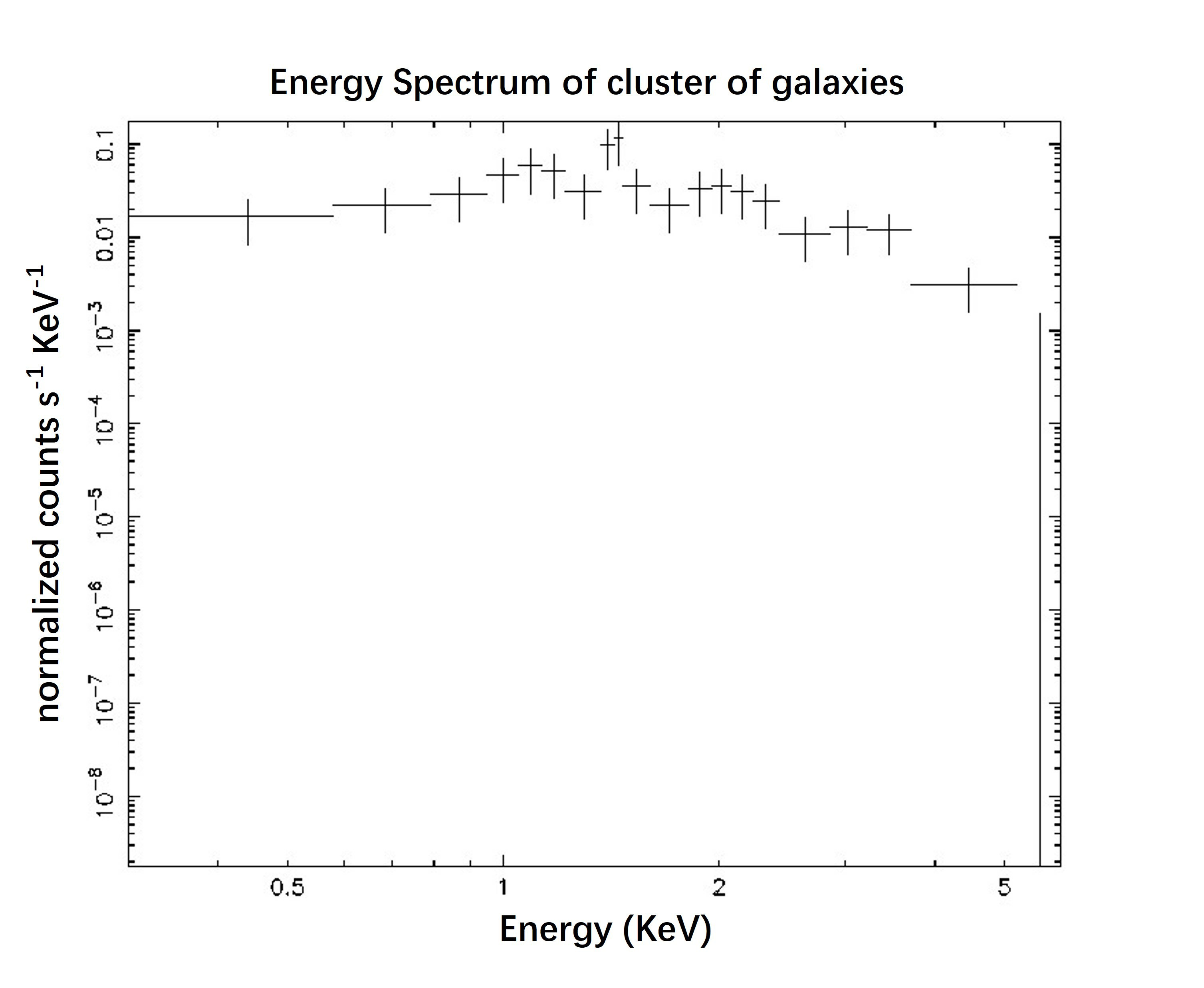}
        \caption{Energy spectrum of cluster of galaxies}
        \label{cluster of galaxies}
    \end{minipage}%
    \begin{minipage}{0.5\textwidth}
        \includegraphics[width=\linewidth]{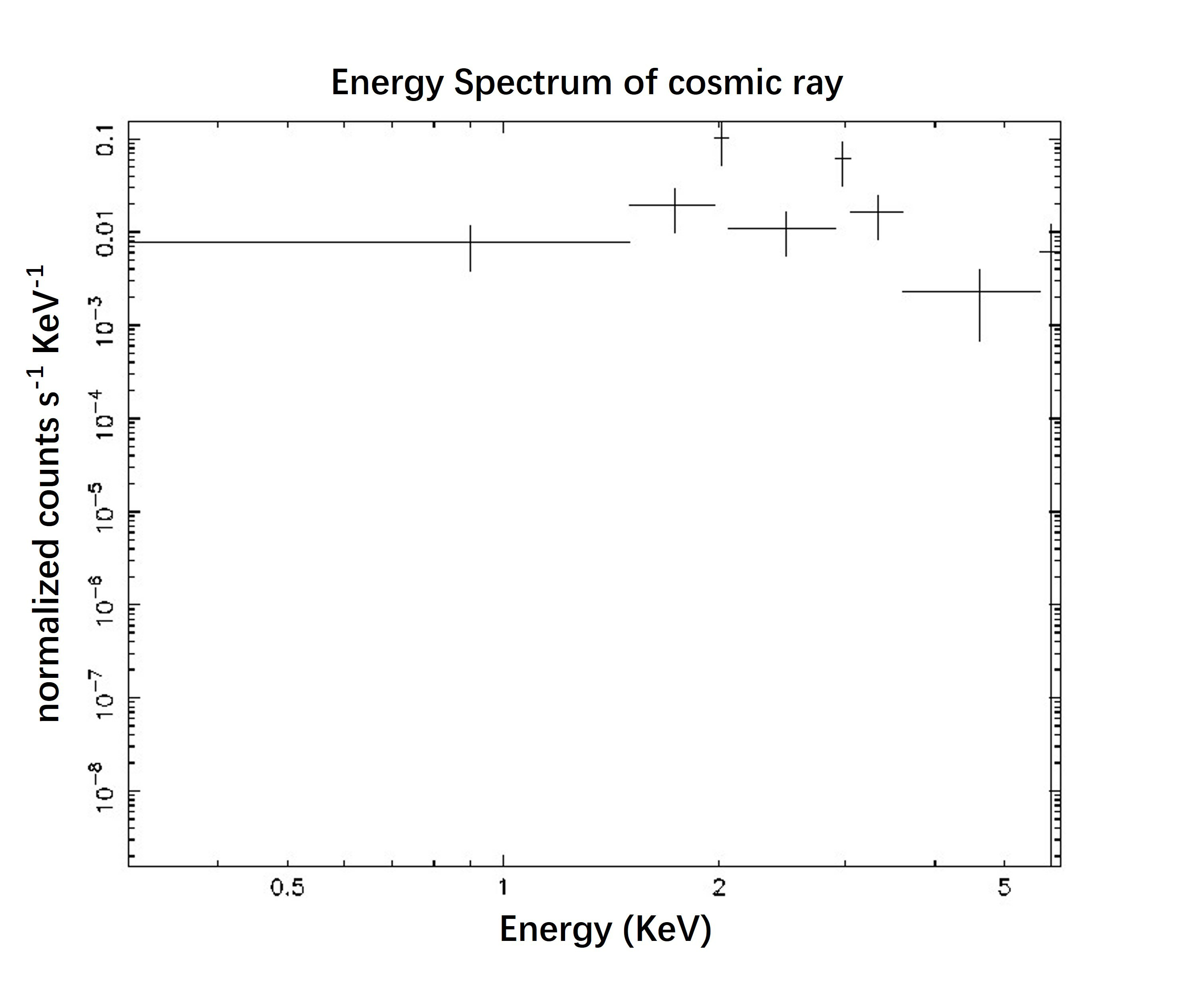}
        \caption{Energy spectrum of cosmic ray}
        \label{cosmic ray}
    \end{minipage}
\end{figure}

\begin{figure}[h!]
    \centering
    \begin{minipage}{0.5\textwidth}
        \includegraphics[width=\linewidth]{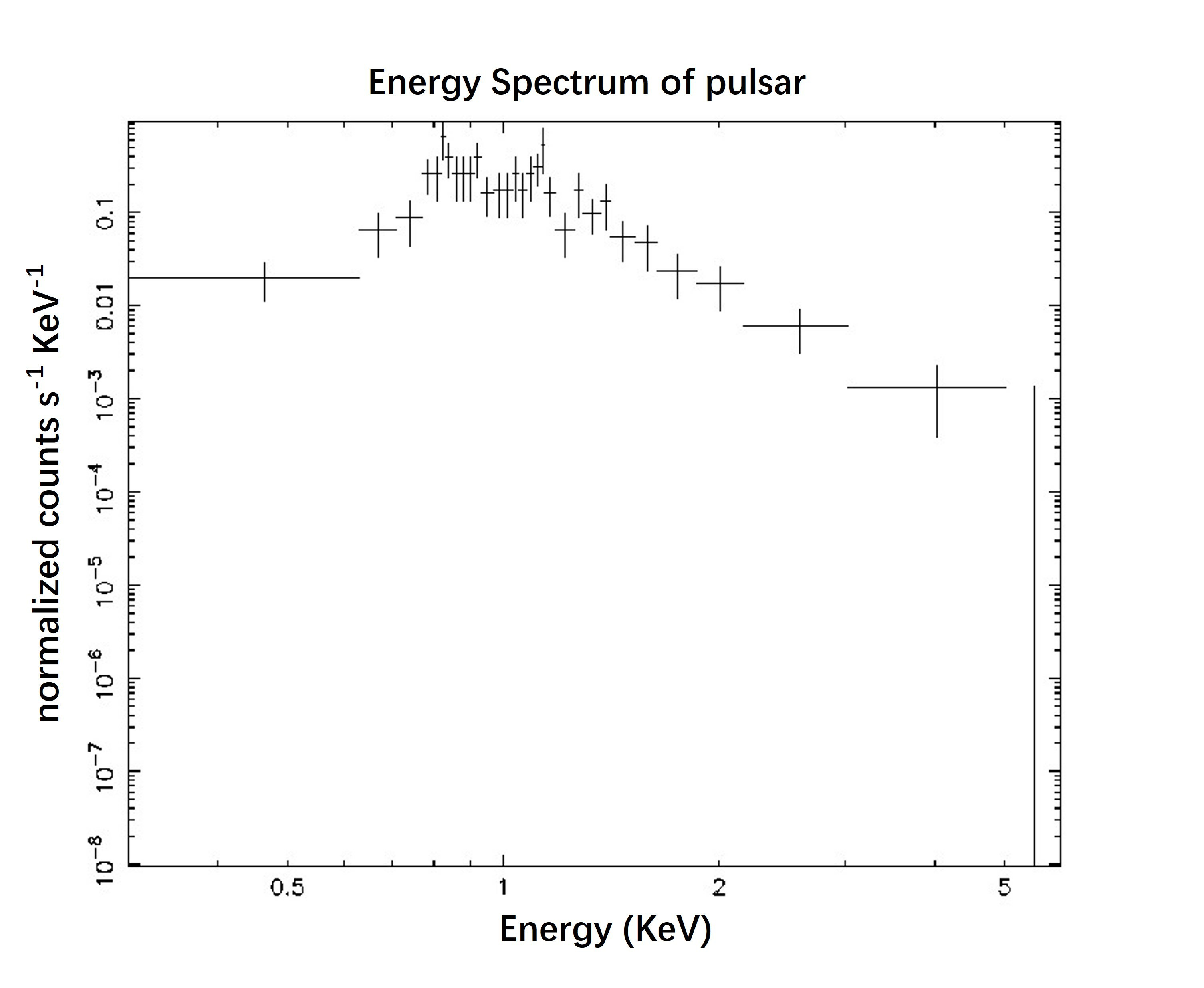}
        \caption{Energy spectrum of pulsar}
        \label{pulsar}
    \end{minipage}%
    \begin{minipage}{0.5\textwidth}
        \includegraphics[width=\linewidth]{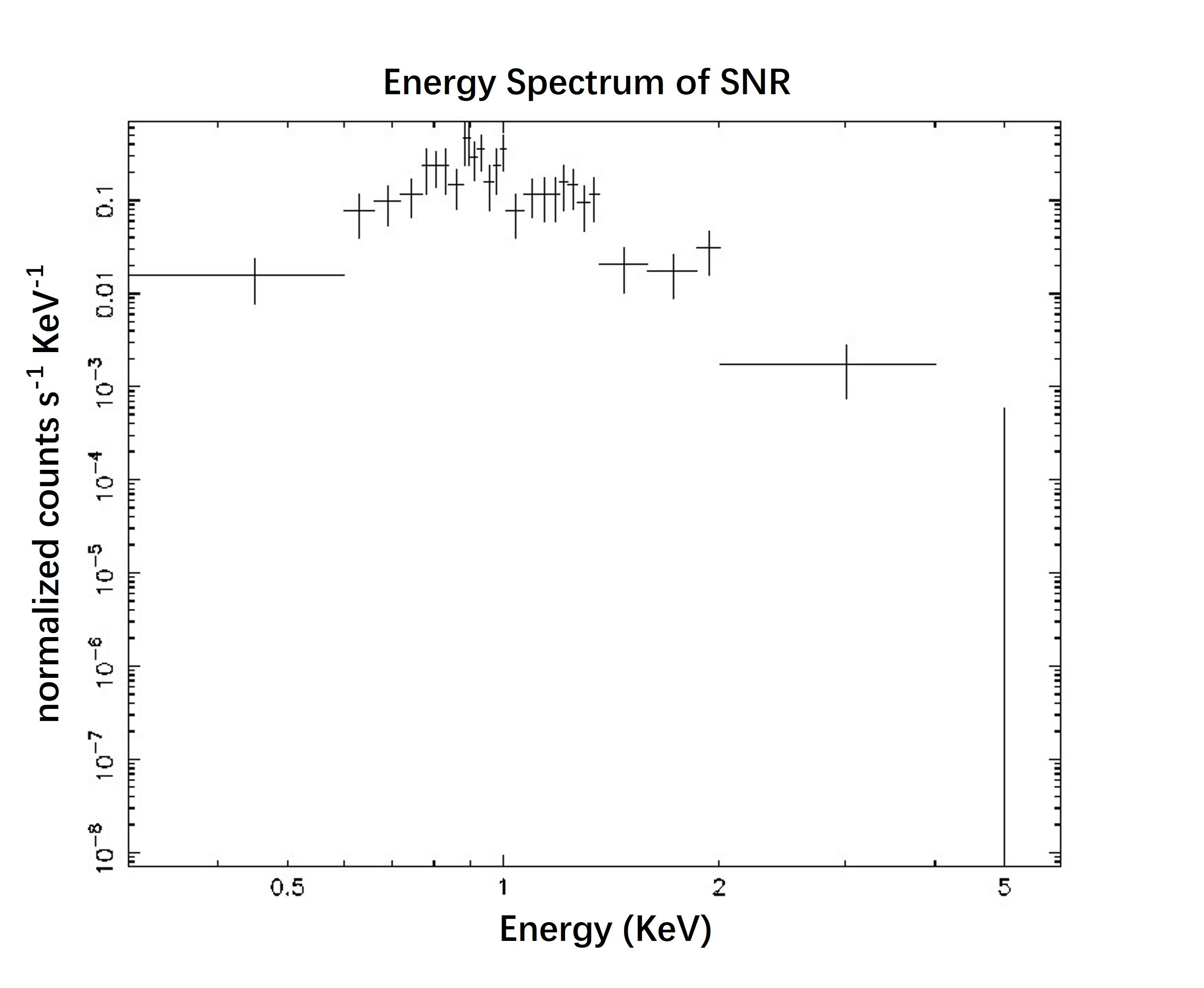}
        \caption{Energy spectrum of SNR}
        \label{SNR}
    \end{minipage}
\end{figure}

\begin{figure}[h!]
    \centering
    \begin{minipage}{0.5\textwidth}
        \includegraphics[width=\linewidth]{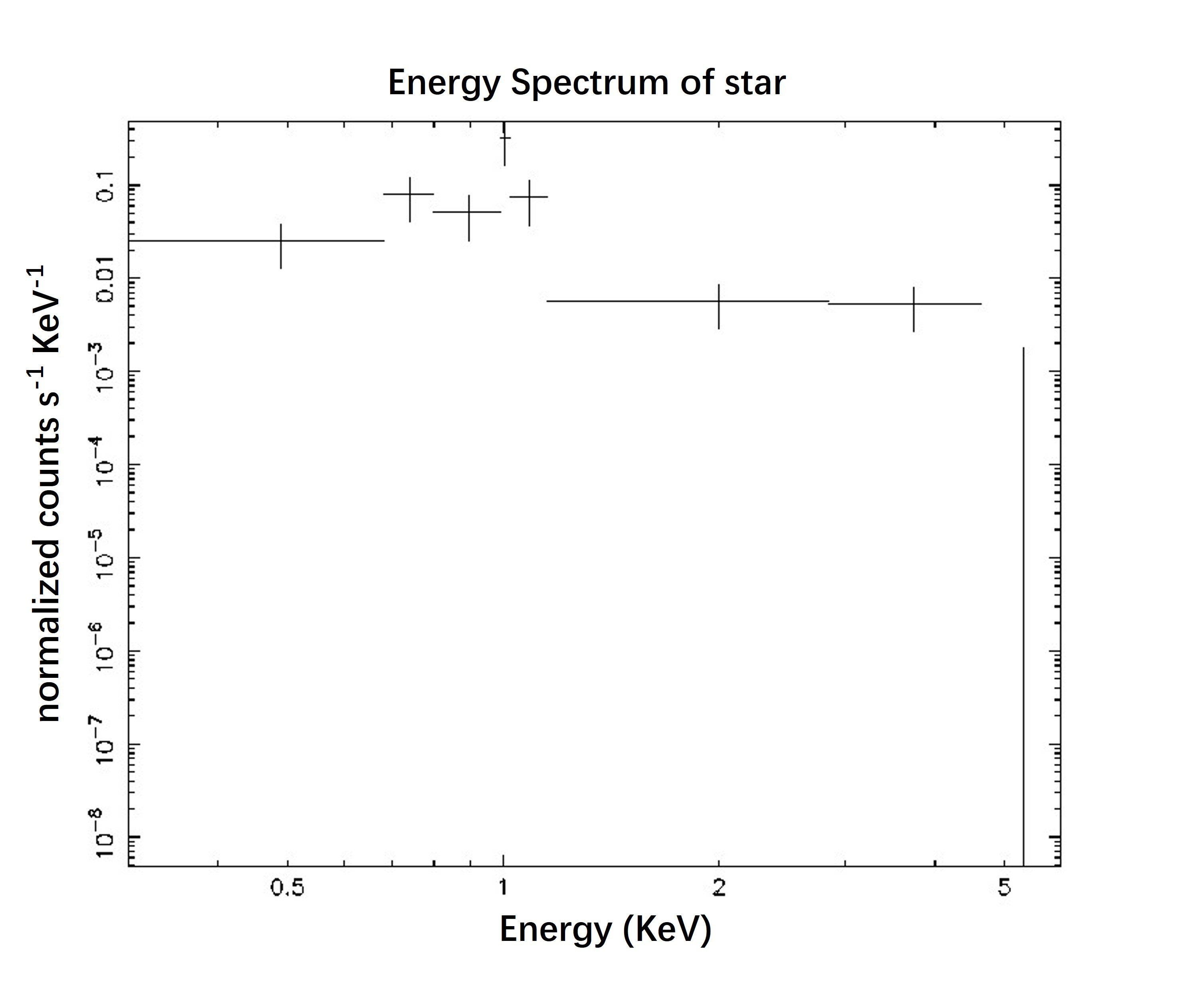}
        \caption{Energy spectrum of star}
        \label{star}
    \end{minipage}%
    \begin{minipage}{0.5\textwidth}
        \includegraphics[width=\linewidth]{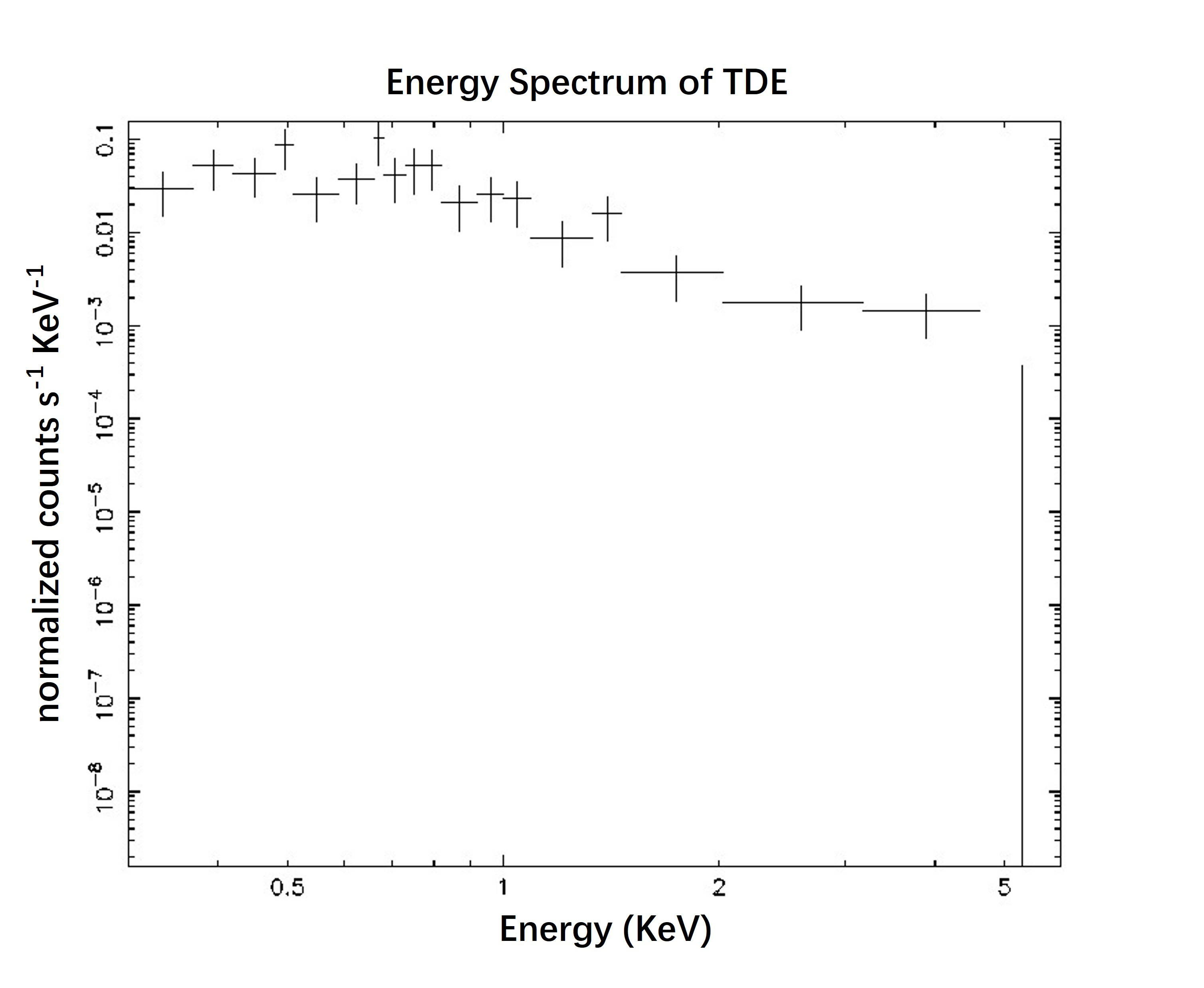}
        \caption{Energy spectrum of TDE}
        \label{TDE}
    \end{minipage}
\end{figure}

\begin{figure}[h!]
    \centering
    \begin{minipage}{0.5\textwidth}
        \includegraphics[width=\linewidth]{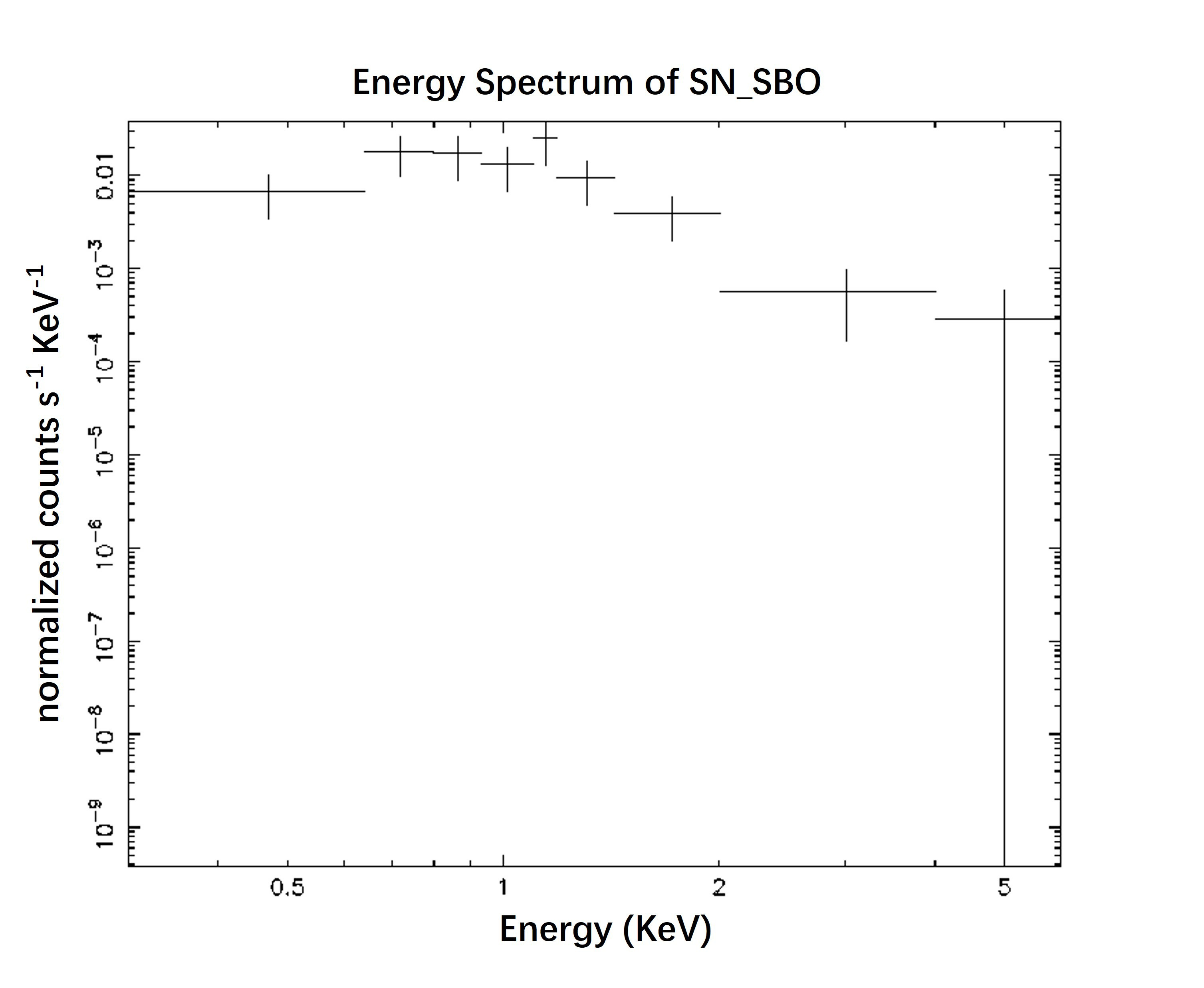}
        \caption{Energy spectrum of SNSBO}
        \label{SNSBO}
    \end{minipage}%
    \begin{minipage}{0.5\textwidth}
        \includegraphics[width=\linewidth]{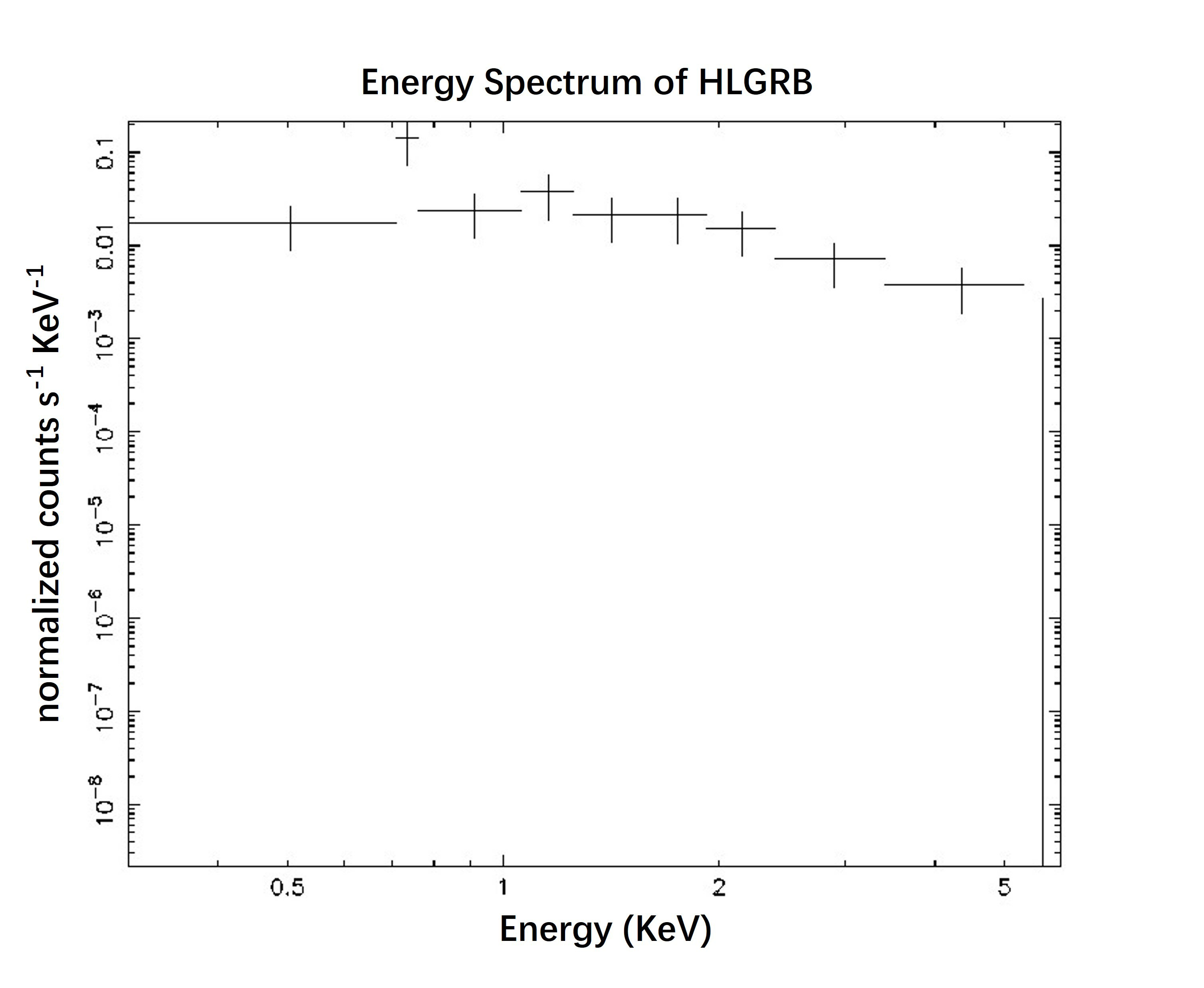}
        \caption{Energy spectrum of HLGRB}
        \label{GRB}
    \end{minipage}
\end{figure}

\clearpage
\section{The light curve for each class}\label{app:lightcurves}

\begin{figure}[h!]
    \centering
    \begin{minipage}{0.5\textwidth}
        \includegraphics[width=\linewidth]{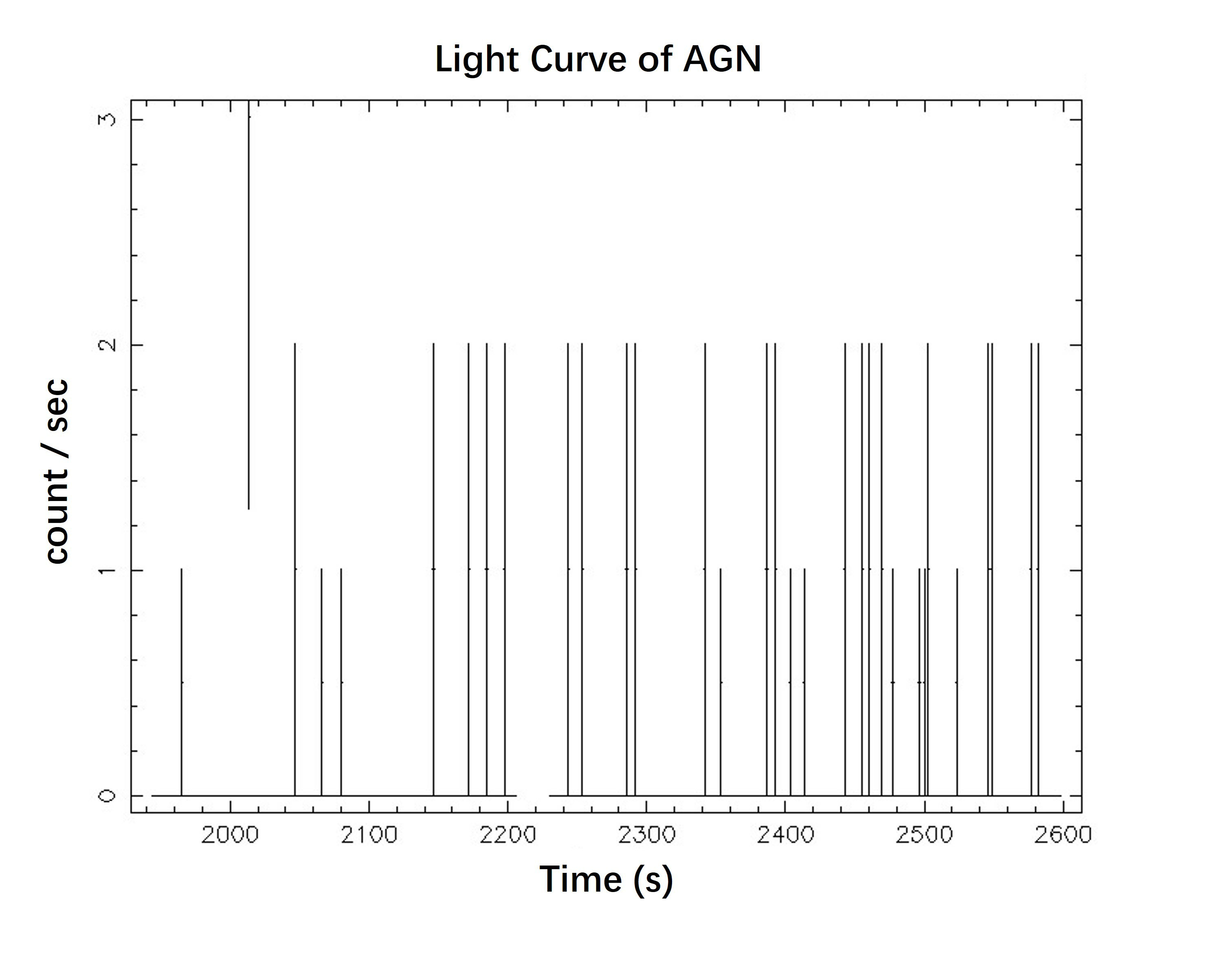}
        \caption{Light curve of AGN}
        \label{AGN LC}
    \end{minipage}%
    \begin{minipage}{0.5\textwidth}
        \includegraphics[width=\linewidth]{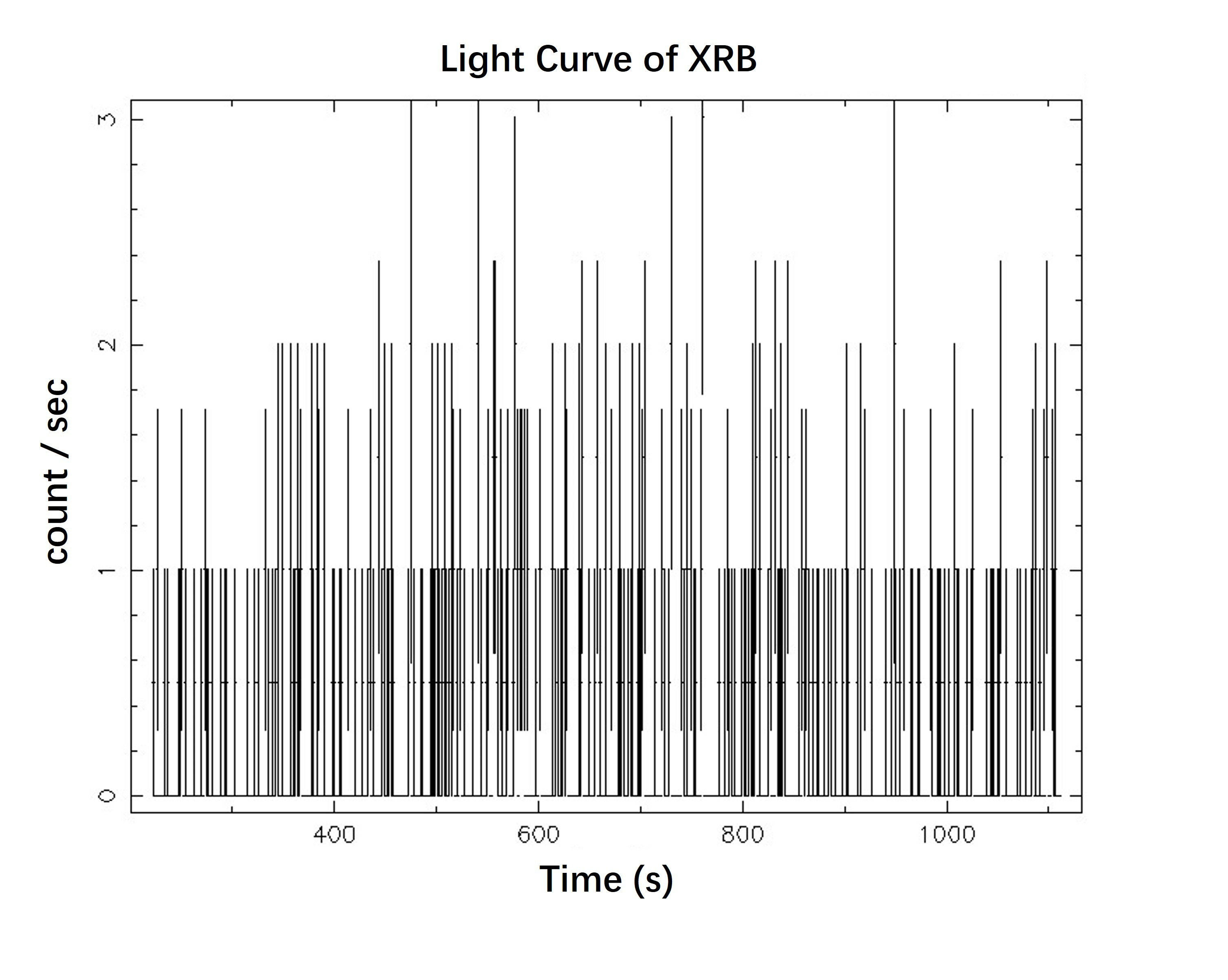}
        \caption{Light curve of XRB}
        \label{XRB LC}
    \end{minipage}
\end{figure}

\begin{figure}[h!]
    \centering
    \begin{minipage}{0.5\textwidth}
        \includegraphics[width=\linewidth]{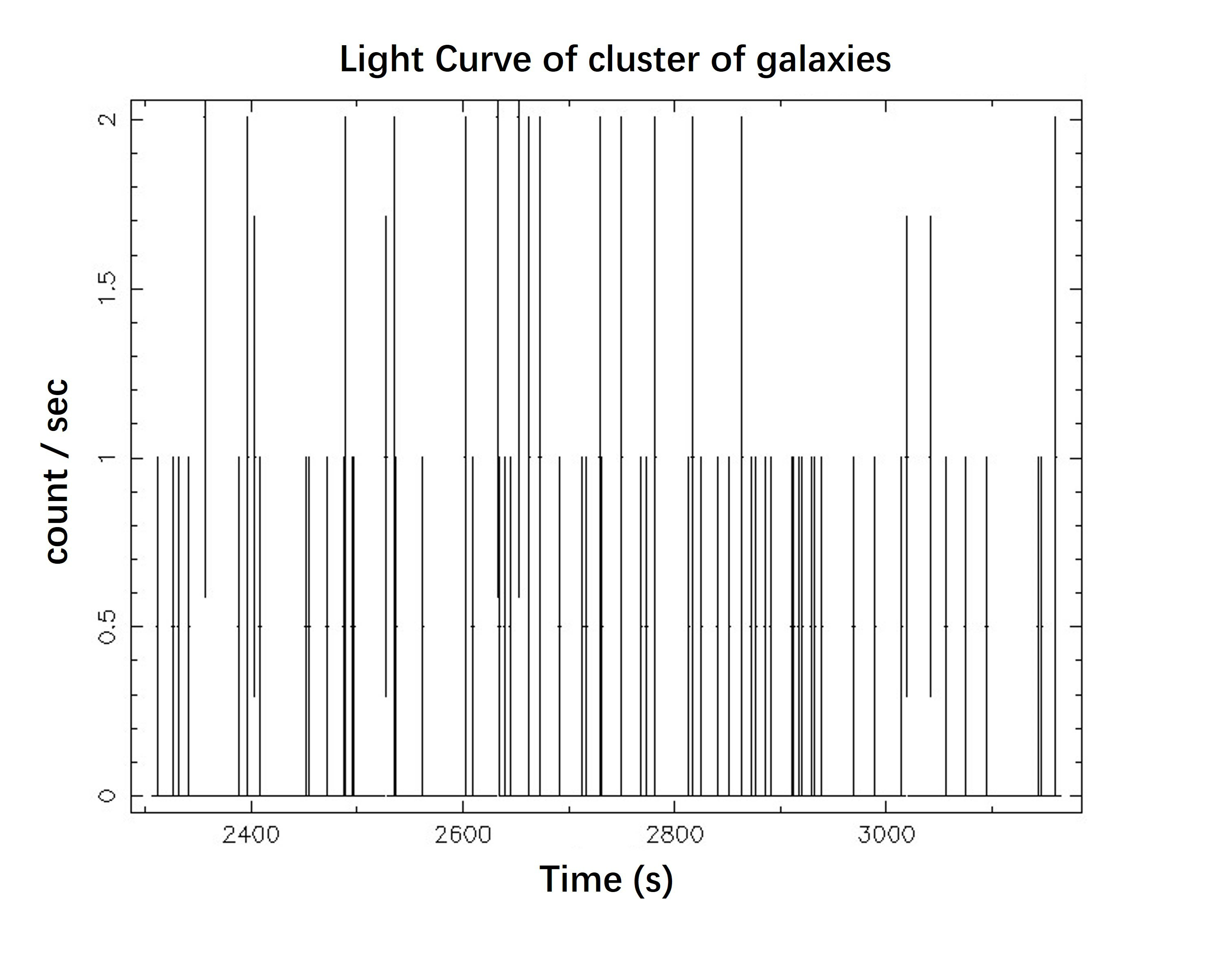}
        \caption{Light curve of cluster of galaxies}
        \label{cluster of galaxies LC}
    \end{minipage}%
    \begin{minipage}{0.5\textwidth}
        \includegraphics[width=\linewidth]{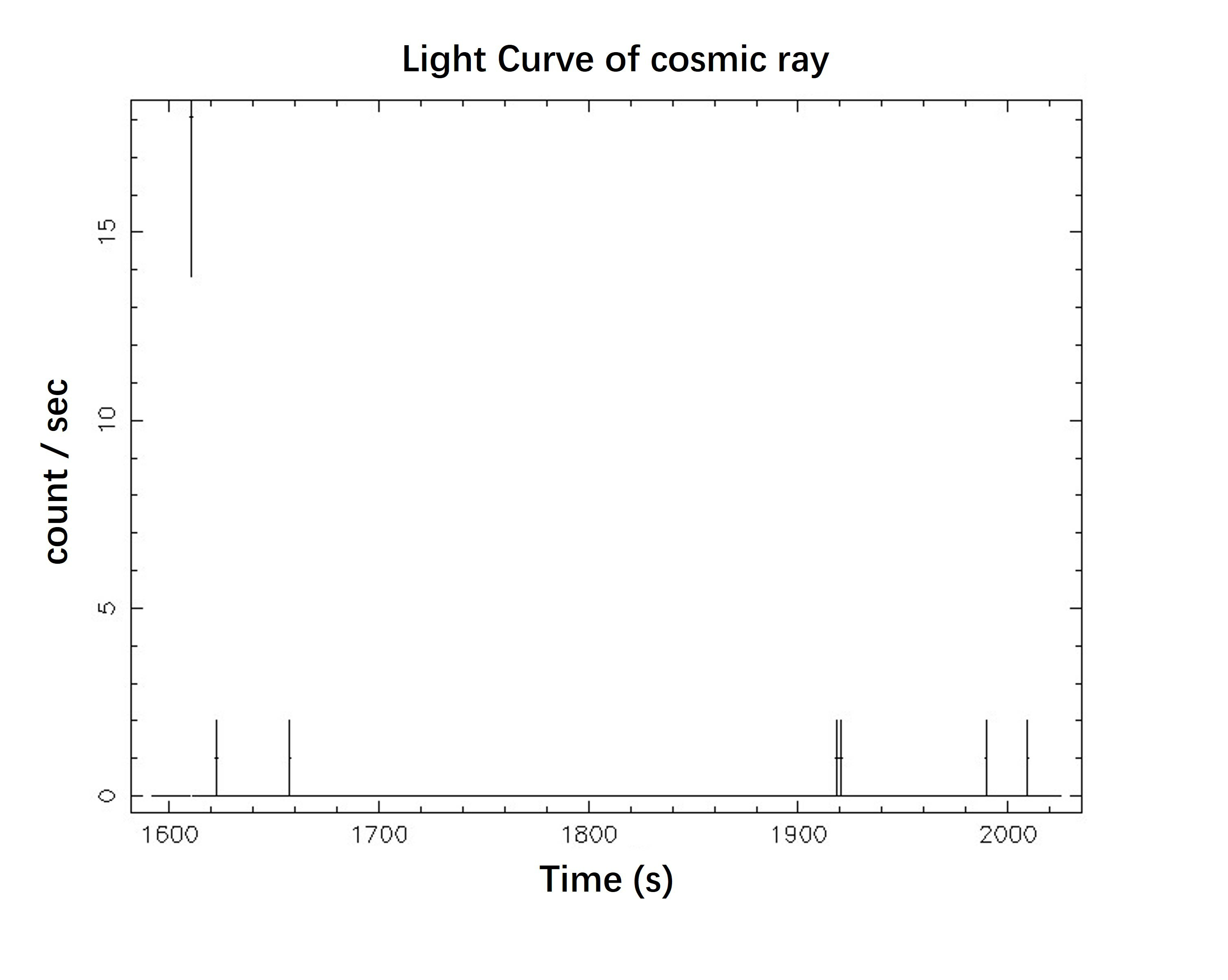}
        \caption{Light curve of cosmic ray}
        \label{cosmic ray LC}
    \end{minipage}
\end{figure}

\begin{figure}[h!]
    \centering
    \begin{minipage}{0.5\textwidth}
        \includegraphics[width=\linewidth]{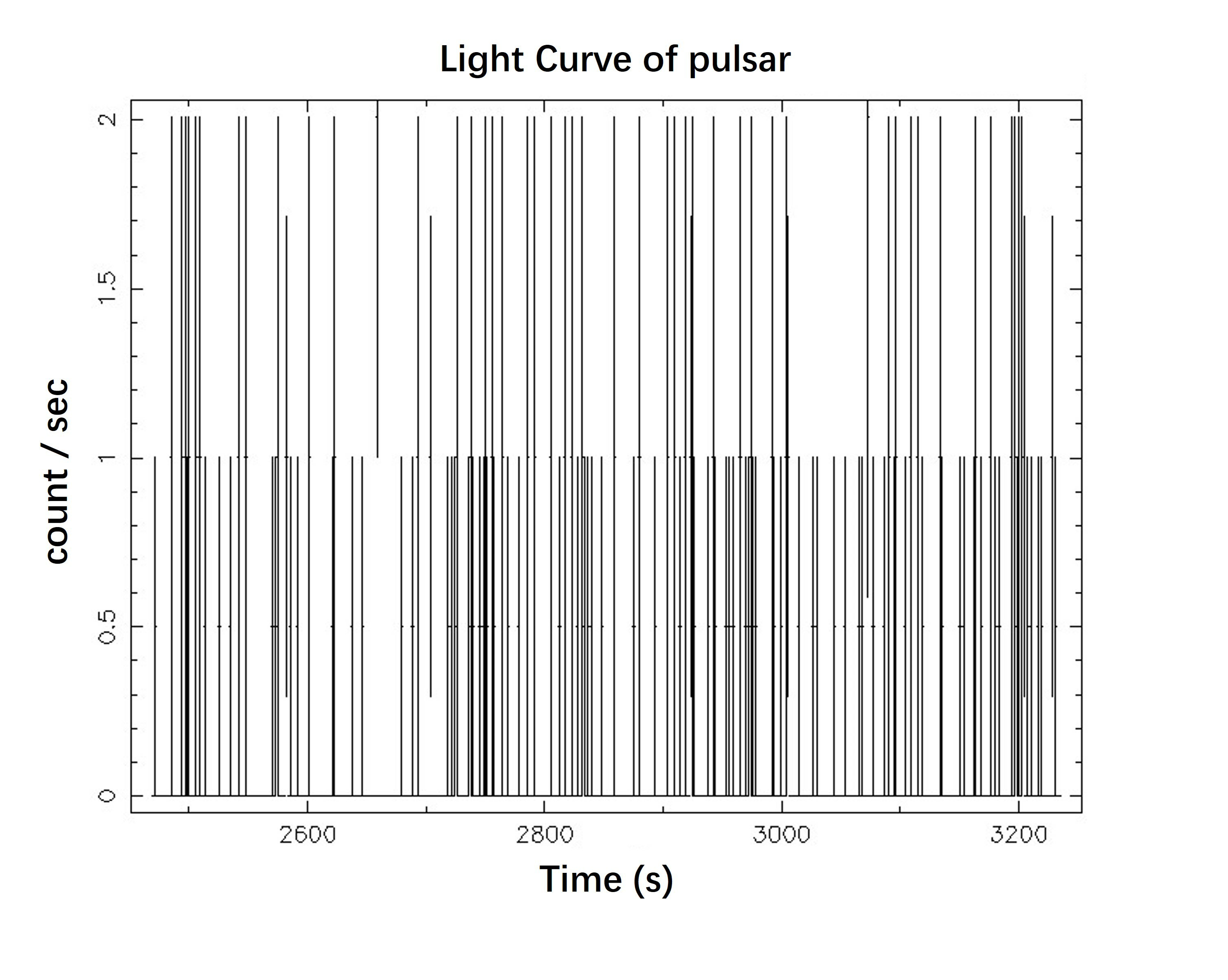}
        \caption{Light curve of pulsar}
        \label{pulsar LC}
    \end{minipage}%
    \begin{minipage}{0.5\textwidth}
        \includegraphics[width=\linewidth]{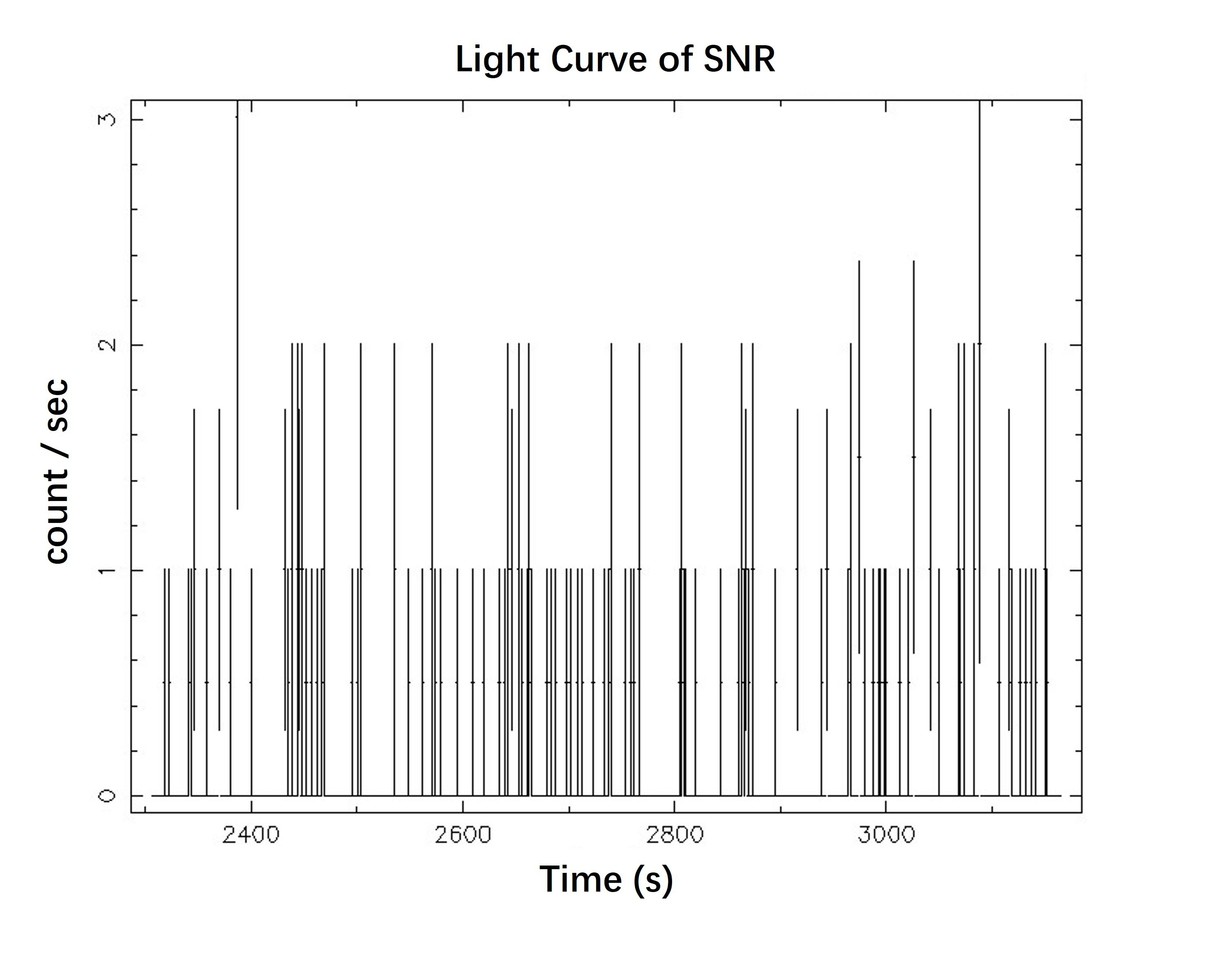}
        \caption{Light curve of SNR}
        \label{SNR LC}
    \end{minipage}
\end{figure}

\begin{figure}[h!]
    \centering
    \begin{minipage}{0.5\textwidth}
        \includegraphics[width=\linewidth]{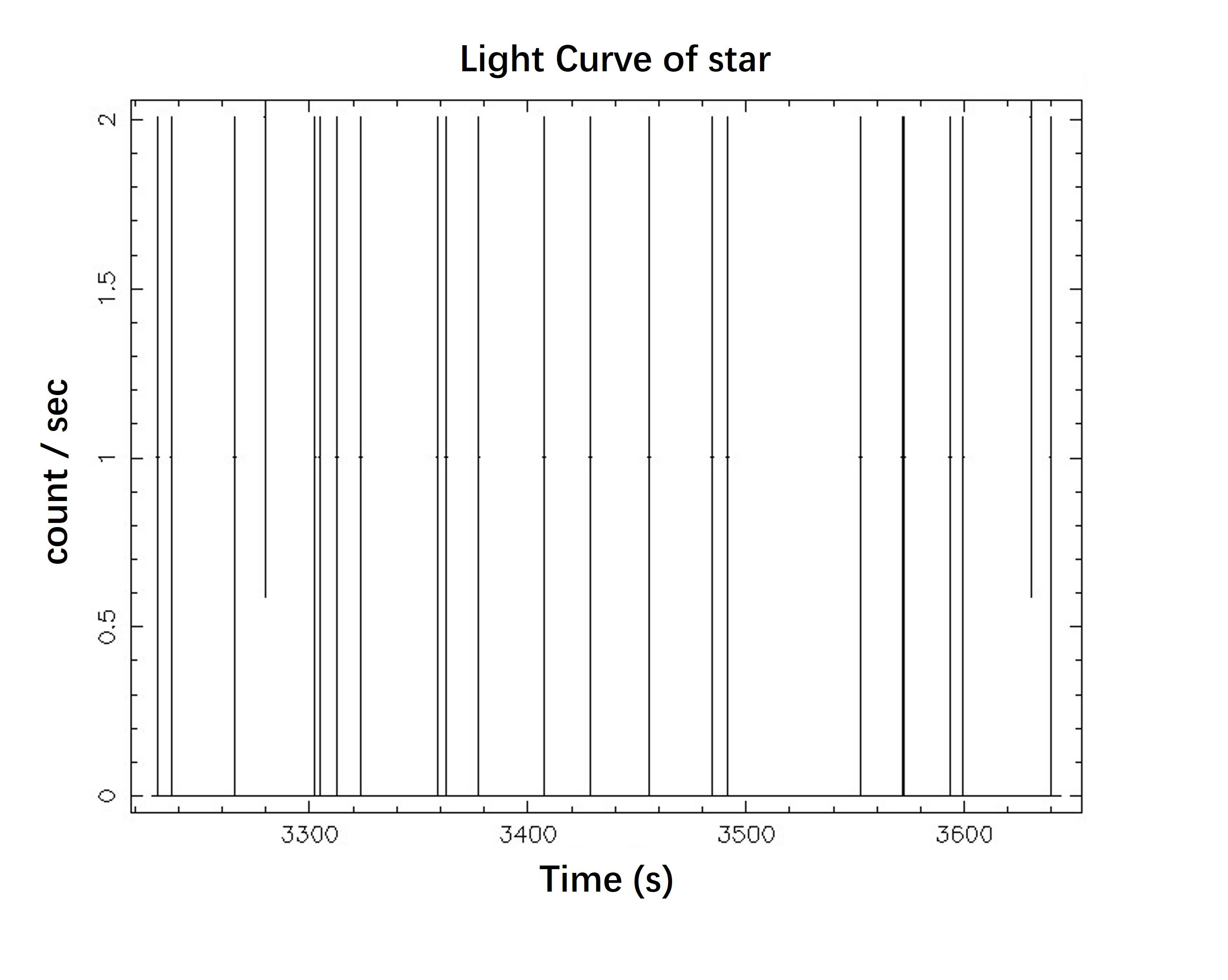}
        \caption{Light curve of star}
        \label{star LC}
    \end{minipage}%
    \begin{minipage}{0.5\textwidth}
        \includegraphics[width=\linewidth]{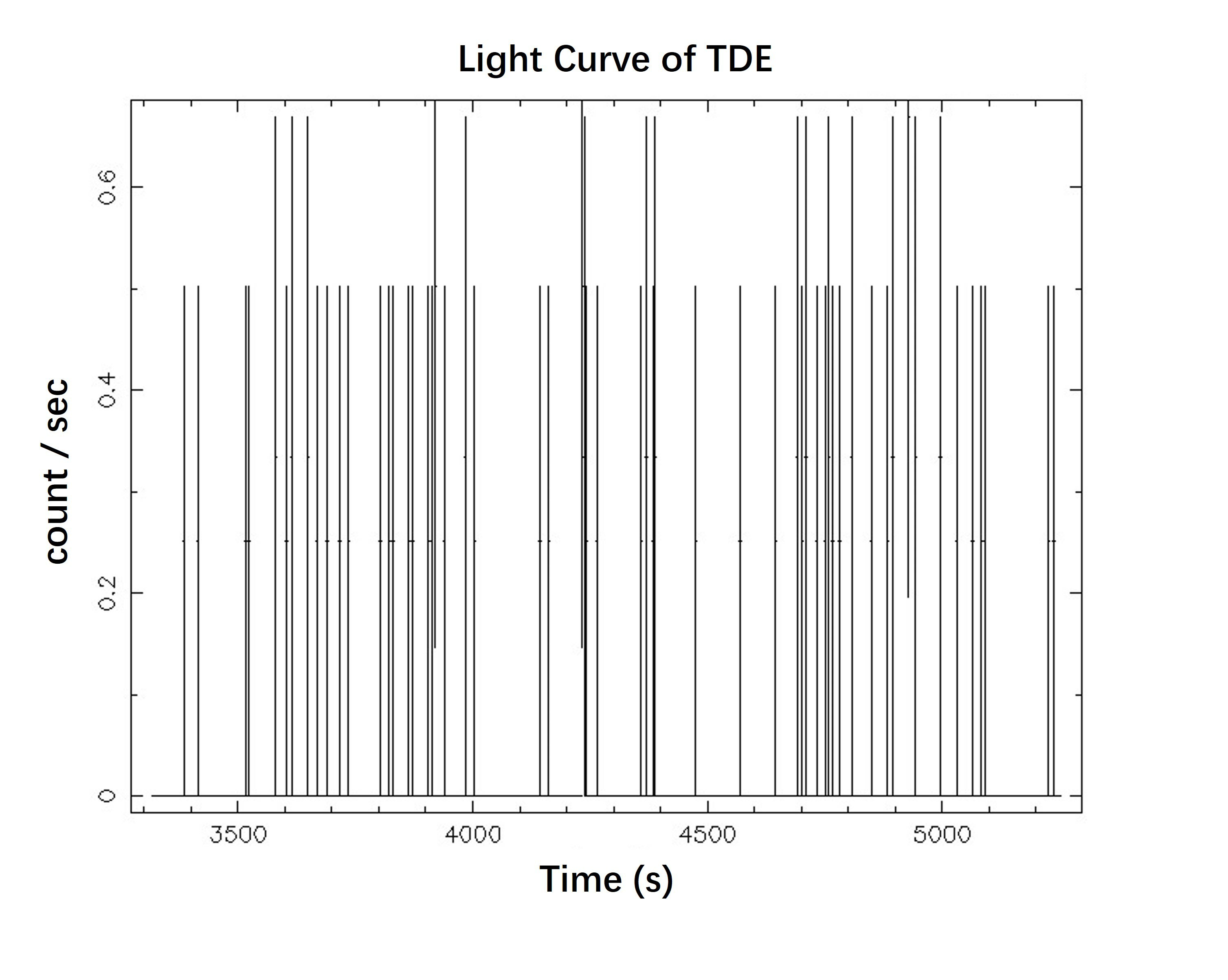}
        \caption{Light curve of TDE}
        \label{TDE LC}
    \end{minipage}
\end{figure}

\begin{figure}[h!]
    \centering
    \begin{minipage}{0.5\textwidth}
        \includegraphics[width=\linewidth]{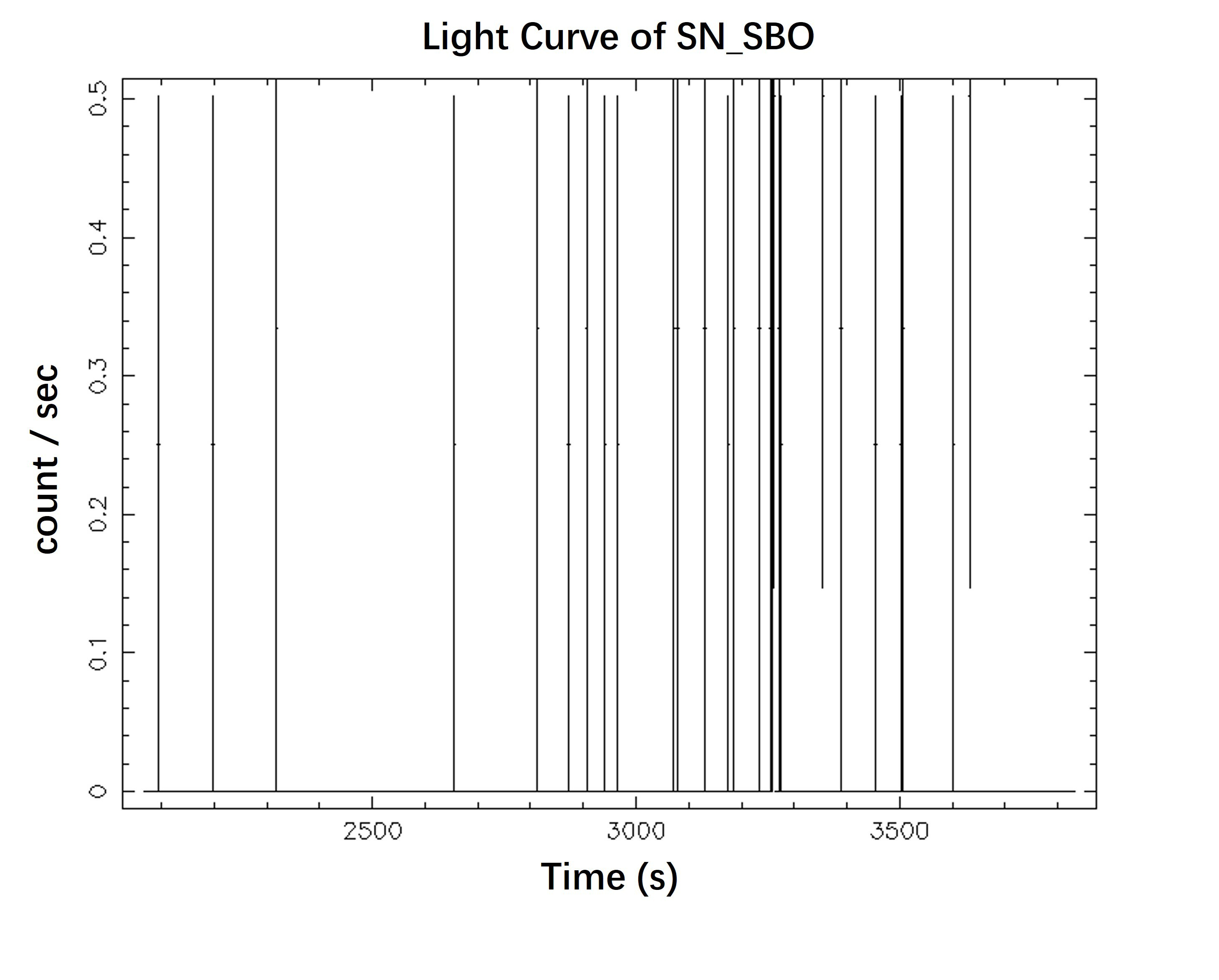}
        \caption{Light curve of SNSBO}
        \label{SNSBO LC}
    \end{minipage}%
    \begin{minipage}{0.5\textwidth}
        \includegraphics[width=\linewidth]{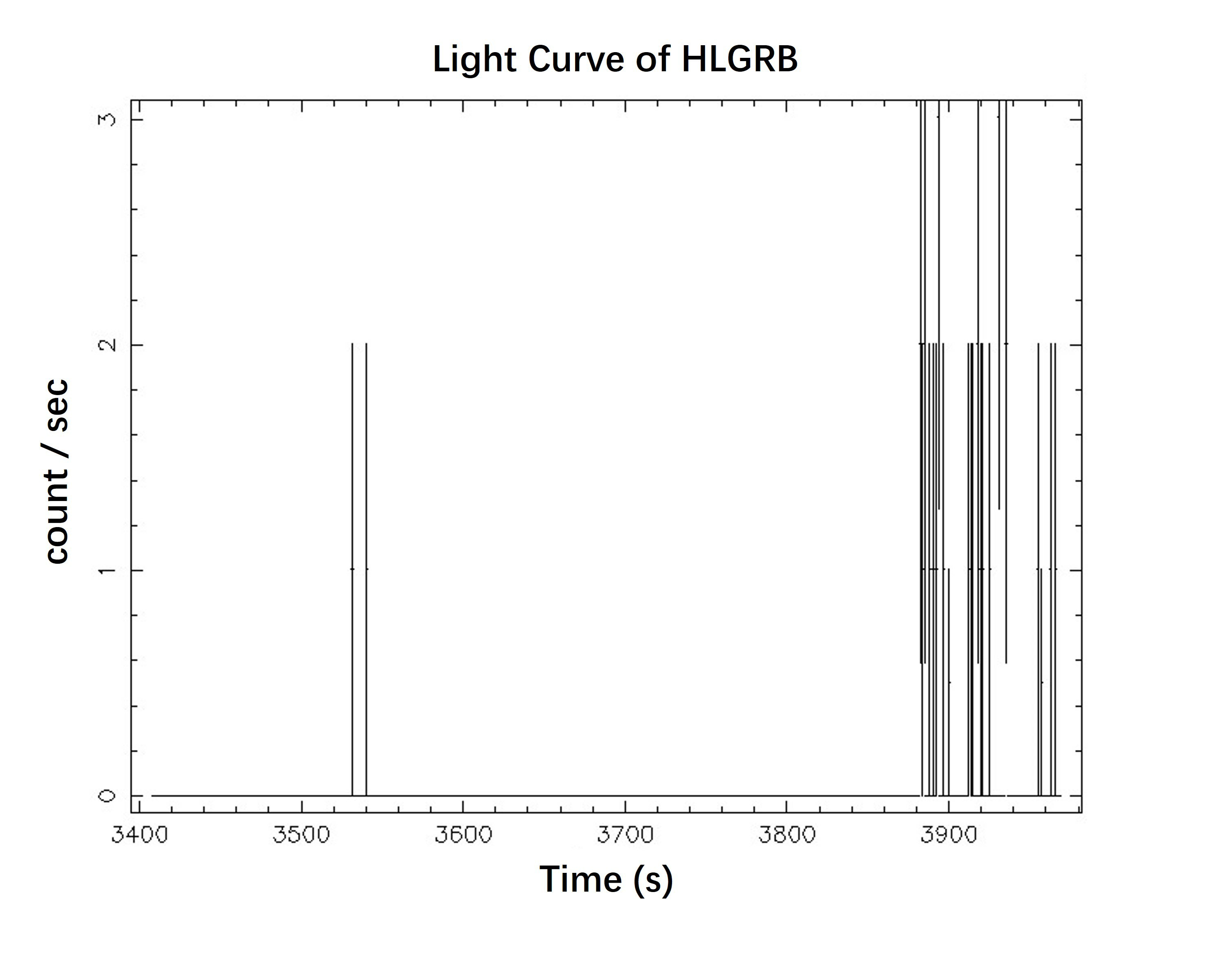}
        \caption{Light curve of HLGRB}
        \label{HLGRB LC}
    \end{minipage}
\end{figure}


\clearpage
\bibliographystyle{raa}
\bibliography{bibtex}

\begin{thebibliography}{30}
\providecommand\natexlab[1]{#1}
\providecommand\JournalTitle[1]{#1}

\bibitem[Angel(1979)]{angel1979lobster}
Angel, J. 1979, in Space Optics Imaging X-Ray Optics Workshop, Vol. 184, SPIE, 84

\bibitem[Breiman(2001)]{breiman2001random}
Breiman, L. 2001, Machine learning, 45, 5

\bibitem[Burrows {et~al.}(2005)]{burrows2005swift}
Burrows, D.~N., Hill, J., Nousek, J., {et~al.} 2005, Space science reviews, 120, 165

\bibitem[Chawla {et~al.}(2002)]{chawla2002smote}
Chawla, N.~V., Bowyer, K.~W., Hall, L.~O., \& Kegelmeyer, W.~P. 2002, Journal of artificial intelligence research, 16, 321

\bibitem[Chen \& Guestrin(2016)]{chen2016xgboost}
Chen, T., \& Guestrin, C. 2016, in Proceedings of the 22nd acm sigkdd international conference on knowledge discovery and data mining, 785

\bibitem[Cortes \& Vapnik(1995)]{cortes1995support}
Cortes, C., \& Vapnik, V. 1995, Machine learning, 20, 273

\bibitem[Cover \& Hart(1967)]{cover1967nearest}
Cover, T., \& Hart, P. 1967, IEEE transactions on information theory, 13, 21

\bibitem[Jansen {et~al.}(2001)]{jansen2001xmm}
Jansen, F., Lumb, D., Altieri, B., {et~al.} 2001, Astronomy \& Astrophysics, 365, L1

\bibitem[Jia {et~al.}(2023)]{jia2023target}
Jia, P., Liu, W., Liu, Y., \& Pan, H. 2023, The Astrophysical Journal Supplement Series, 264, 43

\bibitem[Li {et~al.}(2022)]{li2022swift}
Li, D., Ling, Z., Liu, Y., {et~al.} 2022, The Astronomer's Telegram, 15754, 1

\bibitem[Ling {et~al.}(2022)]{ling2022leia}
Ling, Z., Liu, Y., Zhang, C., {et~al.} 2022, The Astronomer's Telegram, 15748, 1

\bibitem[Lo {et~al.}(2014)]{lo2014automatic}
Lo, K.~K., Farrell, S., Murphy, T., \& Gaensler, B. 2014, The Astrophysical Journal, 786, 20

\bibitem[McGlynn {et~al.}(2004)]{mcglynn2004automated}
McGlynn, T., Suchkov, A., Winter, E., {et~al.} 2004, The Astrophysical Journal, 616, 1284

\bibitem[Murthy {et~al.}(1994)]{murthy1994system}
Murthy, S.~K., Kasif, S., \& Salzberg, S. 1994, Journal of artificial intelligence research, 2, 1

\bibitem[Pedregosa {et~al.}(2011)]{pedregosa2011scikit}
Pedregosa, F., Varoquaux, G., Gramfort, A., {et~al.} 2011, the Journal of machine Learning research, 12, 2825

\bibitem[René(2010)]{Hudec2010Kirkpatrick}
René, H. 2010, X-Ray Optics and Instrumentation, 2010, 1

\bibitem[Richards {et~al.}(2011)]{richards2011machine}
Richards, J.~W., Starr, D.~L., Butler, N.~R., {et~al.} 2011, The Astrophysical Journal, 733, 10

\bibitem[Sun {et~al.}(2023)]{sun2023leia}
Sun, H., Wang, C., Hu, J., {et~al.} 2023, The Astronomer's Telegram, 16195, 1

\bibitem[Tranin {et~al.}(2022)]{tranin2022probabilistic}
Tranin, H., Godet, O., Webb, N., \& Primorac, D. 2022, Astronomy \& Astrophysics, 657, A138

\bibitem[Virtanen {et~al.}(2020)]{2020SciPy-NMeth}
Virtanen, P., Gommers, R., Oliphant, T.~E., {et~al.} 2020, Nature Methods, 17, 261

\bibitem[Voges {et~al.}(1999)]{voges1999rosat}
Voges, W., Aschenbach, B., Boller, T., {et~al.} 1999, arXiv preprint astro-ph/9909315

\bibitem[Weisskopf {et~al.}(2000)]{weisskopf2000chandra}
Weisskopf, M.~C., Tananbaum, H.~D., Van~Speybroeck, L.~P., \& O'Dell, S.~L. 2000, X-ray optics, instruments, and missions iii, 4012, 2

\bibitem[Yang {et~al.}(2022)]{yang2022classifying}
Yang, H., Hare, J., Kargaltsev, O., {et~al.} 2022, The Astrophysical Journal, 941, 104

\bibitem[Yang {et~al.}(2023)]{yang2023leia}
Yang, H., Zhang, W., Jin, C., {et~al.} 2023, The Astronomer's Telegram, 16352, 1

\bibitem[Yuan {et~al.}(2015)]{yuan2015einstein}
Yuan, W., Zhang, C., Feng, H., {et~al.} 2015, arXiv preprint arXiv:1506.07735

\bibitem[Yuan {et~al.}(2018{\natexlab{a}})]{yuan2018einstein}
Yuan, W., Zhang, C., Ling, Z., {et~al.} 2018{\natexlab{a}}, in Space Telescopes and Instrumentation 2018: Ultraviolet to Gamma Ray, Vol. 10699, SPIE, 543

\bibitem[Yuan {et~al.}(2018{\natexlab{b}})]{yuan2018einstein1}
Yuan, W., Zhang, C., Ling, Z., {et~al.} 2018{\natexlab{b}}, Scientia Sinica: Physica, Mechanica et Astronomica, 48, 039502

\bibitem[Zhang {et~al.}(2022{\natexlab{a}})]{zhang2022first}
Zhang, C., Ling, Z., Sun, X., {et~al.} 2022{\natexlab{a}}, The Astrophysical Journal Letters, 941, L2

\bibitem[Zhang {et~al.}(2022{\natexlab{b}})]{zhang2022estimate}
Zhang, J., Qi, L., Yang, Y., {et~al.} 2022{\natexlab{b}}, Astroparticle Physics, 137, 102668

\bibitem[Zhang {et~al.}(2021)]{zhang2021classification}
Zhang, Y., Zhao, Y., \& Wu, X.-B. 2021, Monthly Notices of the Royal Astronomical Society, 503, 5263

\end{thebibliography}

\label{lastpage}

\end{document}